\documentclass[tight,twocolumn]{aastex62}
\usepackage{apjfonts}
\usepackage{float}
\usepackage[caption=false]{subfig}

\usepackage{graphics} 
\usepackage{amssymb,amsmath}
\usepackage{color}

\newcommand{\host}{\hbox{2MASX J07001137$-$6602251}}

\def\lsim{\lower0.3em\hbox{$\,\buildrel <\over\sim\,$}}
\def\gsim{\lower0.3em\hbox{$\,\buildrel >\over\sim\,$}}

\newcommand{\msun}{\hbox{M$_{\odot}$}}

\newcommand{\halpha}{\hbox{H$\alpha$}}
\newcommand{\hbeta}{\hbox{H$\beta$}}

\newcommand{\swift}{\textit{Swift}}

\begin{document}

\title{Discovery and Early Evolution of ASASSN-19bt, the First TDE Detected by TESS}
\shorttitle{Discovery of the TDE ASASSN-19bt} 
\shortauthors{Holoien, et al. }

\author[0000-0001-9206-3460]{Thomas~W.-S.~Holoien}
\altaffiliation{Carnegie Fellow}
\affiliation{The Observatories of the Carnegie Institution for Science, 813 Santa Barbara St., Pasadena, CA 91101, USA}

\author{Patrick~J.~Vallely}
\altaffiliation{NSF Graduate Fellow}
\affiliation{Department of Astronomy, The Ohio State University, 140 West 18th Avenue, Columbus, OH 43210, USA}

\author{Katie~Auchettl}
\affiliation{DARK, Niels Bohr Institute, University of Copenhagen, Lyngbyvej 2, 2100 Copenhagen, Denmark}

\author{K.~Z.~Stanek}
\affiliation{Department of Astronomy, The Ohio State University, 140 West 18th Avenue, Columbus, OH 43210, USA}
\affiliation{Center for Cosmology and AstroParticle Physics (CCAPP), The Ohio State University, 191 W.\ Woodruff Ave., Columbus, OH 43210, USA}

\author{Christopher~S.~Kochanek}
\altaffiliation{Radcliffe Fellow}
\affiliation{Department of Astronomy, The Ohio State University, 140 West 18th Avenue, Columbus, OH 43210, USA}
\affiliation{Center for Cosmology and AstroParticle Physics (CCAPP), The Ohio State University, 191 W.\ Woodruff Ave., Columbus, OH 43210, USA}

\author[0000-0002-4235-7337]{K.~Decker~French}
\altaffiliation{Hubble Fellow}
\affiliation{The Observatories of the Carnegie Institution for Science, 813 Santa Barbara St., Pasadena, CA 91101, USA}

\author{Jose~L.~Prieto}
\affiliation{N\'ucleo de Astronom\'ia de la Facultad de Ingenier\'ia y Ciencias, Universidad Diego Portales, Av. Ej\'ercito 441, Santiago, Chile}
\affiliation{Millennium Institute of Astrophysics, Santiago, Chile}

\author{Benjamin~J.~Shappee}
\affiliation{Institute for Astronomy, University of Hawai'i, 2680 Woodlawn Drive, Honolulu, HI 96822, USA}

\author{Jonathan~S.~Brown}
\affiliation{Department of Astronomy and Astrophysics, University of California, Santa Cruz, CA 95064, USA}

\author{Michael~M.~Fausnaugh}
\affiliation{Department of Physics and Kavli Institute for Astrophysics and Space Research, Massachusetts Institute of Technology, Cambridge, MA 02139, USA}

\author{Subo~Dong}
\affiliation{Kavli Institute for Astronomy and Astrophysics, Peking University, Yi He Yuan Road 5, Hai Dian District, Beijing 100871, China}

\author{Todd~A.~Thompson}
\affiliation{Department of Astronomy, The Ohio State University, 140 West 18th Avenue, Columbus, OH 43210, USA}
\affiliation{Center for Cosmology and AstroParticle Physics (CCAPP), The Ohio State University, 191 W.\ Woodruff Ave., Columbus, OH 43210, USA}
\affiliation{Institute for Advanced Study, 1 Einstein Dr., Princeton, NJ 08540, USA}

\author{Subhash~Bose}
\affiliation{Kavli Institute for Astronomy and Astrophysics, Peking University, Yi He Yuan Road 5, Hai Dian District, Beijing 100871, China}

\author{Jack~M.~M.~Neustadt}
\affiliation{Department of Astronomy, The Ohio State University, 140 West 18th Avenue, Columbus, OH 43210, USA}

\author{P.~Cacella}
\affiliation{DogsHeaven Observatory, SMPW Q25 CJ1 LT10B, Brasilia, DF 71745-501, Brazil}

\author{J.~Brimacombe}
\affiliation{Coral Towers Observatory, Cairns, Queensland 4870, Australia}

\author{Malhar~R.~Kendurkar}
\affiliation{Prince George Astronomical Observatory, 7765 Tedford Road, Prince George, British Columbia, V2N 6S2, Canada}

\author[0000-0002-1691-8217]{Rachael~L.~Beaton}
\altaffiliation{Hubble Fellow}
\altaffiliation{Carnegie-Princeton Fellow}
\affiliation{Department of Astrophysical Sciences, Princeton University, 4 Ivy Lane, Princeton, NJ~08544, USA}
\affiliation{The Observatories of the Carnegie Institution for Science, 813 Santa Barbara St., Pasadena, CA 91101, USA}

\author{Konstantina~Boutsia}
\affiliation{Las Campanas Observatory, Carnegie Observatories, Casilla 601, La Serena, Chille.}

\author{Laura~Chomiuk}
\affiliation{Center for Data Intensive and Time Domain Astronomy, Department of Physics and Astronomy, Michigan State University, East Lansing, MI 48824, USA}

\author[0000-0002-7898-7664]{Thomas~Connor}
\affiliation{The Observatories of the Carnegie Institution for Science, 813 Santa Barbara St., Pasadena, CA 91101, USA}

\author{Nidia~Morrell}
\affiliation{Las Campanas Observatory, Carnegie Observatories, Casilla 601, La Serena, Chille.}

\author{Andrew~B.~Newman}
\affiliation{The Observatories of the Carnegie Institution for Science, 813 Santa Barbara St., Pasadena, CA 91101, USA}

\author{Gwen~C.~Rudie}
\affiliation{The Observatories of the Carnegie Institution for Science, 813 Santa Barbara St., Pasadena, CA 91101, USA}

\author{Laura~Shishkovksy}
\affiliation{Center for Data Intensive and Time Domain Astronomy, Department of Physics and Astronomy, Michigan State University, East Lansing, MI 48824, USA}

\author{Jay~Strader}
\affiliation{Center for Data Intensive and Time Domain Astronomy, Department of Physics and Astronomy, Michigan State University, East Lansing, MI 48824, USA}

\correspondingauthor{T.~W.-S.~Holoien}
\email{tholoien@carnegiescience.edu}

\date{\today}

\begin{abstract}
We present the discovery and early evolution of ASASSN-19bt, a tidal disruption event (TDE) discovered by the All-Sky Automated Survey for Supernovae (ASAS-SN) at a distance of $d\simeq115$~Mpc and the first TDE to be detected by TESS. As the TDE is located in the TESS Continuous Viewing Zone, our dataset includes 30-minute cadence observations starting on 2018 July 25, and we precisely measure that the TDE begins to brighten $\sim8.3$ days before its discovery. Our dataset also includes 18 epochs of {\swift} UVOT and XRT observations, 2 epochs of {\emph XMM-Newton} observations, 13 spectroscopic observations, and ground data from the Las Cumbres Observatory telescope network, spanning from 32 days before peak through 37 days after peak. ASASSN-19bt thus has the most detailed pre-peak dataset for any TDE. The TESS light curve indicates that the transient began to brighten on 2019 January 21.6 and that for the first 15 days its rise was consistent with a flux $\propto t^2$ power-law model. The optical/UV emission is well-fit by a blackbody SED, and ASASSN-19bt exhibits an early spike in its luminosity and temperature roughly 32 rest-frame days before peak and spanning up to 14 days that has not been seen in other TDEs, possibly because UV observations were not triggered early enough to detect it. It peaked on 2019 March 04.9 at a luminosity of $L\simeq1.3\times10^{44}$~ergs~s$^{-1}$ and radiated $E\simeq3.2\times10^{50}$~ergs during the 41-day rise to peak. X-ray observations after peak indicate a softening of the hard X-ray emission prior to peak, reminiscent of the hard/soft states in X-ray binaries.
\end{abstract}
\keywords{accretion, accretion disks --- black hole physics --- galaxies: nuclei}


\section{Introduction}
\label{sec:intro}

Tidal disruption events (TDEs) are rare transient phenomena that occur when a star passes within the tidal radius of a supermassive black hole (SMBH). This results in the tidal forces from the SMBH overwhelming the self-gravity of the star, tearing the star apart. In the classical picture, roughly half of the disrupted stellar material is ejected from the system while the rest remains bound to the SMBH and falls back to pericenter at a rate asymptotically proportional to $t^{-5/3}$. A portion of this material then forms an accretion disk, producing a luminous, short-lived flare \citep[e.g.,][]{lacy82,rees88,evans89,phinney89}. 

The initial theoretical work \citep[e.g.,][]{lacy82,rees88,evans89,phinney89} predicted temperatures of $T\sim$ a few $\times10^5$~K, which would result in emission peaking at soft X-ray energies, and that the transient luminosity would follow the same $t^{-5/3}$ time dependence as the mass fallback rate. In recent years, however, wide-area sky surveys have discovered an increasing number of TDE candidates that exhibit a large range of observational properties that differ from the classical picture \citep[e.g.,][]{velzen11,cenko12a,gezari12b,arcavi14, chornock14,holoien14b,gezari15,vinko15,holoien16a,holoien16b,brown16a,auchettl17,blagorodnova17,brown17a,brown17b,gezari17,holoien18a,holoien18b,velzen19,leloudas19}. These objects typically exhibit nearly constant temperatures roughly an order of magnitude cooler than the initial predictions, peaking at ultraviolet (UV) wavelengths; a wide range of luminosity decline rates that vary over time; and broad hydrogen and helium emission lines of varying relative strength in their optical spectra. Despite significant theoretical work on these objects, a single, unifying model has yet to be developed that can explain all the observations. However, it is now clear that TDE emission depends on a range of factors, including the properties of the disrupted star \citep[e.g.,][]{macleod12,kochanek16}, the evolution of the stellar debris stream after disruption \citep[e.g.,][] {kochanek94,strubbe09,guillochon13,hayasaki13,hayasaki16,piran15,shiokawa15}, and radiative transfer effects \citep[e.g.,][]{gaskell14,strubbe15,roth16,roth18}. 

Only a small subset of TDEs have been discovered prior to peak light, making it difficult to study the evolution of the stellar debris following disruption and the formation of the accretion disk, and resulting in a significant gap in our theoretical understanding of these objects. Only in recent years have early discoveries become more common, as sky surveys with high cadence and wide coverage, such as the All-Sky Automated Survey for Supernovae \citep[ASAS-SN;][]{shappee14,kochanek17}, the Palomar Transient Factory \citep[PTF;][]{law09}, the Asteroid Terrestrial-impact Last Alert System \citep[ATLAS;][]{tonry18}, and the Zwicky Transient Facility \citep[ZTF;][]{bellm19}, have grown and come online. This has resulted in a growing sample of TDEs with several weeks of observations prior to peak \citep[e.g.,][]{holoien18a,leloudas19,velzen19,wevers19}, but there has yet to be a case where the transient has been caught within hours of beginning to brighten, as has been done with several supernovae (SNe).

The \emph{Kepler} spacecraft, which continuously monitored thousands of galaxies between its original four-year primary mission and the \emph{K2} Campaign 16 Supernova Experiment, created a new paradigm for early-time SN light curves, as it obtained light curves of six SNe spanning from prior to explosion through the early rise at extremely high cadence. The \emph{Kepler} sample includes ASASSN-18bt, which has the most precisely measured SN light curve to- date \citep{olling15,garnavich16,shappee19,dimitriadis19}. 

The Transiting Exoplanet Survey Satellite \citep[TESS;][]{ricker15} has the potential to do the same for TDEs, as it combines continuous space-based monitoring for time spans ranging from one month to one year with an extremely wide field-of-view, providing both the ability to observe TDEs minutes after they start to brighten and the sky coverage needed to have any likelihood of observing a TDE. TESS has already detected significantly more SNe than \emph{Kepler} in less than a year of operations \citep[][]{fausnaugh19,vallely19}, and can achieve a $3\sigma$ limiting magnitude of $\sim20$ mag in 8 hours of observation, making it an ideal complement to modern high-cadence ground-based surveys.

Here we present the discovery and early-time observations of ASASSN-19bt, a TDE discovered by ASAS-SN on 2019 January 29 in the galaxy \host. ASASSN-19bt is located in the TESS Continuous Viewing Zone (CVZ), and is the first TDE flare detected by TESS, providing us with an unprecedented cadence on its rising light curve and the ability to precisely measure the time when the transient began to brighten. At a redshift of $z=0.0262$ ($d=115.2$~Mpc for $H_0=69.6$~km~s$^{-1}$~Mpc$^{-1}$, $\Omega_M=0.29$, and $\Omega_{\Lambda}=0.71$) based on an archival 6dF spectrum obtained from NED \citep{jones09}, it is also one of the nearest TDEs discovered to-date, and had a peak UV magnitude comparable to that of ASASSN-14li \citep{holoien16a}. In Section~\ref{sec:obs} we describe the discovery of ASASSN-19bt, the available archival data for the host, and the observations obtained in our follow-up campaign. In Section~\ref{sec:analysis} we fit the physical properties of the TDE using the early light curves, describe the transient's blackbody evolution and compare it to other TDEs in literature, and analyze the early spectroscopic evolution of ASASSN-19bt. Finally, a summary of our results and a discussion of the physical implications are given in Section~\ref{sec:disc}.


\section{Discovery and Observations}
\label{sec:obs}

ASAS-SN is an ongoing project designed to monitor the entire visible sky in an unbiased way with a rapid cadence to discover bright, nearby transients. To accomplish this, we use units hosted by the Las Cumbres Observatory global telescope network \citep{brown13} at multiple sites around the globe, each consisting of four 14-cm telescopes on a common mount. ASAS-SN expanded in 2017 and currently operates with five units, located in Hawaii, Texas, Chile, and South Africa. ASAS-SN can observe the entire visible sky to a depth of $g\sim18.5$~mag roughly once every 24 hours, weather permitting \citep{shappee14, kochanek17}. In order to maximize the synergy between ASAS-SN and TESS, ASAS-SN monitors the TESS fields at an increased cadence, allowing us to discover transients in the TESS fields and trigger follow-up data collection as soon as possible in order to complement the TESS light curve. 


\begin{figure*}
\includegraphics[width=\textwidth]{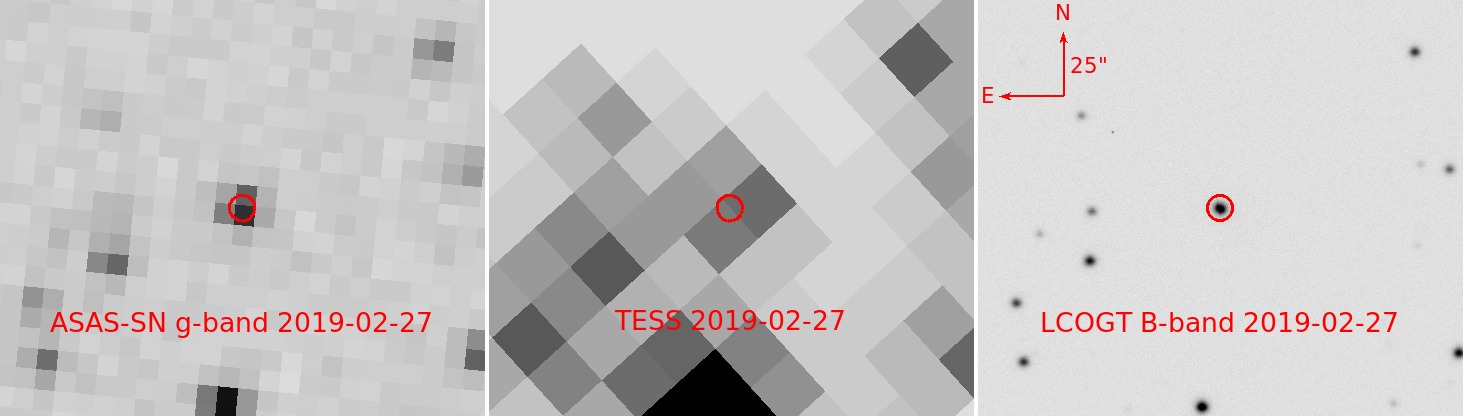}
\caption{Images of ASASSN-19bt near peak light obtained with ASAS-SN (left), TESS (center), and a Las Cumbres Observatory 1-m telescope (right). The red circle has a radius of 5\farcs{0} and is centered on the position of the transient.}
\label{fig:three_images}
\end{figure*}

ASASSN-19bt was discovered on 2019 January 29.91 at  RA$=$07:00:11.41, Dec$=-$66:02:25.16 (J2000) in $g$-band images obtained from the ASAS-SN ``Payne-Gaposchkin'' unit located in Sutherland, South Africa, and we promptly announced its discovery to the community via the Transient Name Server, which assigned it the designation AT 2019ahk\footnote{\url{https://wis-tns.weizmann.ac.il/object/2019ahk}}. Because its position was consistent with the nucleus of its host galaxy and it was located in a field that was part of the TESS CVZ, we triggered a spectroscopic follow-up observation with the Low-Dispersion Survey Spectrograph 3 (LDSS-3) mounted on the Magellan Clay 6.5-m telescope on 2019 January 31.20. The spectrum revealed a strong blue continuum, narrow emission features consistent with those in an archival spectrum of the host (see Section~\ref{sec:archival}), and no obvious broad emission features. Based on this, we publicly announced the discovery and classification of the target as a possible young TDE \citep{cacella19}, noting that a lack of broad hydrogen and helium emission lines prior to peak light has been seen in other TDEs  \citep[e.g.,][]{holoien18a}. Figure~\ref{fig:three_images} shows images of ASASSN-19bt taken near peak light from TESS, ASAS-SN, and the Las Cumbres Observatory 1-m telescopes. Despite the large pixel sizes of ASAS-SN and TESS, the transient is clearly detected by both instruments.

Based on the preliminary classification, we requested and were awarded target-of-opportunity (TOO) observations from the \textit{Neil Gehrels Swift Gamma-ray Burst Mission} \citep[\swift;][]{gehrels04} UltraViolet and Optical Telescope \citep[UVOT;][]{roming05} and X-ray Telescope \citep[XRT;][]{burrows05}. The {\swift} observations confirmed that the transient was UV-bright and exhibited very faint X-ray emission, both consistent with the TDE classification, and we began a multi-wavelength follow-up campaign to observe the transient.


\subsection{Archival Data of \host}
\label{sec:archival}

We obtained $BVgri$ magnitudes of the host galaxy from the AAVSO Photometric All-Sky Survey Data Release 10 \citep[APASS;][]{henden15}, $JHK_S$ magnitudes from the Two Micron All-Sky Survey (2MASS), and $W1$ and $W2$ measurements from the Wide-field Infrared Survey Explorer \citep[WISE;][]{wright10} AllWISE data release. \host{} is at a declination too far south to be observed by optical surveys such as the Sloan Digital Sky Survey (SDSS) and Pan-STARRS or radio surveys like FIRST and NVSS. It is not detected in archival data from, or was not previously observed by, Spitzer, Herschel, or the Hubble Space Telescope (HST). 

Serendipitously, \host{} is located approximately 8\farcm{5} away from the BL Lac-type object PKS~0700$-$661, which was observed several times by {\swift} in 2009, 2010, 2014 and 2018 (target IDs 38456, 41619, and 83377). Due to the 17\farcm{0}$\times$17\farcm{0} field-of-view of {\swift}, \host{} was captured in several of these archival observations. While it is located at the edge of the field, making it difficult to use these archival images as image subtraction templates, we obtained archival \swift{} UVOT magnitudes by first summing all the available data in each of the six UVOT filters using the HEAsoft software task {\tt uvotimsum}, and then extracting counts from the combined images in a 5\farcs{0} radius aperture using the software task {\tt uvotsource}, with a sky aperture of $\sim$~40\farcs{0} radius to estimate and subtract the background counts. We converted the archival count rates to magnitudes and fluxes using the most recent UVOT calibration \citep{poole08,breeveld10}. The {\swift} UVOT, APASS, 2MASS and WISE magnitudes of the host that we used to fit the host galaxy SED and estimate the host flux (see below) are shown in Table~\ref{tab:host_mags}.

There were pre-event X-ray observations of 2MASX J07001137$-$6602251 by the  ROSAT All-Sky Survey \citep{voges99} circa 1990 and in the serendipitous {\swift} X-Ray Telescope \citep[XRT;][]{burrows05} observations. No source was detected by ROSAT to a 3$\sigma$ upper limit of $4.7\times10^{-3}$ counts s$^{-1}$ in the 0.3-2.0 keV range. Assuming a $\Gamma=1.75$ power-law spectrum typical of an AGN \citep[e.g.,][]{ricci17}, and an HI column density of 7.1$\times10^{20}$ cm$^{-2}$, we derive a 3$\sigma$ absorbed flux upper limit of $\sim 1.7\times10^{-12}$ ergs cm$^{-2}$~s$^{-1}$ over the 0.3-10.0 keV energy range. This corresponds to an absorbed luminosity of $2.7\times10^{42}$ erg s$^{-1}$, implying that the host does not harbour a strong AGN \citep[e.g.,][]{ricci17}. Interestingly, we detect weak ($\sim3\sigma)$ X-ray emission from the first {\swift} observation (ID~38450) in 2009 but a  follow-up observation (also ID~38450) taken $\sim20$ days later does not show any significant X-ray emission. Further observations (IDs 41619 and 83377) taken in 2010, 2014 and 2018 show no evidence of X-ray emission. 

The detection in the first {\swift} observation has a count rate of  ($2.3\pm1)\times10^{-3}$ counts/sec in the 0.3-10.0 keV energy range. Assuming the same model we used to derive the ROSAT upper limit, this corresponds to a flux of  $(9.6\pm4)\times10^{-14}$ erg cm$^{-2}$ s$^{-1}$ and a luminosity of $(1.5\pm0.7)\times10^{41}$ erg s$^{-1}$ in the 0.3-10.0 keV energy range. This emission could be indicative of a weak AGN, but there are too few counts to determine the origin of the emission.

If we then combine all the {\swift} non-detections, we obtain a $3\sigma$ upper limit on the count rate of $9.7\times10^{-4}$, which corresponds to a flux limit of $4.1\times10^{-14}$ erg cm$^{-2}$ s$^{-1}$ and a luminosity limit of $6.3\times10^{40}$ erg s$^{-1}$. For our BH mass estimate (see below), the apparent detection corresponds to $ 2\times 10^{-4}$ of Eddington given our estimated BH mass, and the upper limit from the {\swift} non-detections is two times lower.


\begin{deluxetable}{ccc}
\tabletypesize{\footnotesize}
\tablecaption{Archival Host Photometry}
\tablehead{
\colhead{Filter} &
\colhead{Magnitude} &
\colhead{Magnitude Uncertainty} }
\startdata
$UVW2$ & 19.55 & 0.06 \\
$UVM2$ & 19.59 & 0.07 \\
$UVW1$ & 19.08 & 0.06 \\
$U_{UVOT}$ & 18.10 & 0.04 \\
$B_{J}$ & 16.69 & 0.05 \\
$g$ & 16.38 & 0.06 \\
$V_J$ & 15.98 & 0.01 \\
$r$ & 15.73 & 0.06 \\
$i$ & 15.30 & 0.10 \\ 
$J$ & 14.55 & 0.05 \\
$H$ & 14.26 & 0.06 \\
$K_S$ & 14.47 & 0.11 \\
$W1$ & 15.33 & 0.02 \\
$W2$ & 16.00 & 0.02 
\enddata 
\tablecomments{Archival magnitudes of {\host} from {\swift} (UV$+U$), APASS ($BgVri$), 2MASS ($JHK_S$), and WISE ($W1$, $W2$) used as inputs for host SED fitting with \textsc{fast}. Magnitudes in the {\swift} filters are 5\farcs{0} aperture magnitudes measured from archival data, while APASS, 2MASS, and WISE magnitudes were taken from their respective catalogs. The {\swift} and APASS magnitudes shown here were also used to calculate the host's flux in the UV and optical bands to use for host subtraction from our follow-up photometry (see Sections \ref{sec:swift} and \ref{sec:other_obs}). All magnitudes are  in the AB system.} 
\label{tab:host_mags} 
\end{deluxetable}

\begin{figure*}
\begin{minipage}{\textwidth}
\centering
\subfloat{\includegraphics[width=0.48\textwidth]{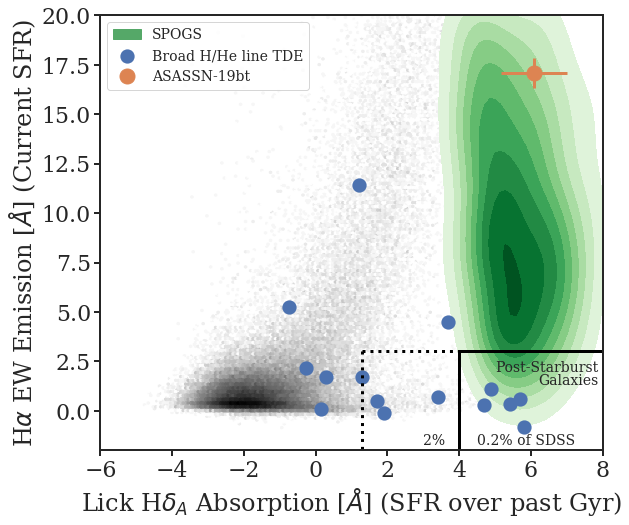}}
\subfloat{\includegraphics[width=0.48\textwidth]{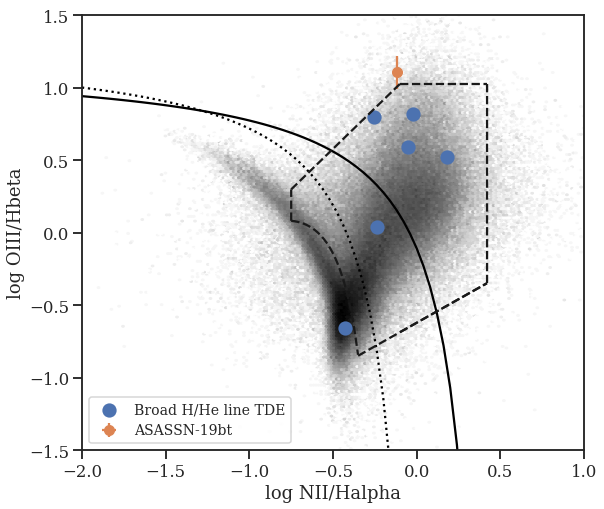}}
\caption{\emph{Left Panel}: Lick H$\delta_A$ absorption (tracing star formation over the last Gyr) compared to the H$\alpha$ EW emission (usually tracing current star formation) for {\host} in orange and several other TDE hosts in blue. Galaxies from the SDSS main spectroscopic survey \citep{strauss02} are shown in grey. The green contours show the population of shocked post-starburst (SPOG) galaxies \citep{alatalo16}. {\host} has strong H $\alpha$ emission like many star-forming galaxies, but its strong Balmer absorption places it closer to the population of SPOGs. \emph{Right Panel}: BPT diagram showing {\host} (orange) and other TDE host galaxies (blue).  Galaxies from the SDSS main spectroscopic survey \citep{strauss02} are shown in grey. The solid line is the theoretical maximum starburst line from \cite{kewley01} and the dotted line is the observed starburst - AGN separation from \cite{kauffmann03}. The region enclosed by dashed lines indicates galaxies with line ratios consistent with shocks. {\host} is in the Seyfert or non-star-forming portion of this diagram but slightly outside of the shocked region, so it is similar to, but not a part of, the SPOG population.}
\label{fig:host_lines}
\end{minipage}
\end{figure*}

 We assessed the recent star formation history by measuring the H$\alpha$ equivalent width (EW) and Lick H$\delta_A$ index of the 6dF spectrum as in \cite{french16}, finding that the H$\alpha$ EW is $17.11 \pm 0.76$ \AA\ and the Lick H$\delta_A$ index is $6.08 \pm 0.92$ \AA, where we are defining positive values of EW as emission. The left panel of Figure~\ref{fig:host_lines} shows the H$\alpha$ EW compared to the Lick H$\delta_A$ index for {\host} and several other TDE host galaxies. While its Lick H$\delta_A$ index is similar to those of other TDE hosts, {\host} exhibits a significantly stronger H$\alpha$ emission feature. We also measured the NII6584/H$\alpha$ and OIII5007/H$\beta$ line ratios in order to analyze the galaxy using the BPT diagram \citep{baldwin81} shown in the right panel of Figure~\ref{fig:host_lines}. The NII6584/H$\alpha$ flux ratio is $0.76 \pm 0.03$ and the  OIII5007/H$\beta$ flux ratio is $12.85 \pm 3.36$. These line ratios indicate the possible presence of a Seyfert in the host galaxy, but could also be generated by shocks \citep[e.g.,][]{rich15}. 


As can be seen in the left panel of Figure~\ref{fig:host_lines}, 2MASX J07001137-6602251 is similar to the "shocked post-starburst" (SPOG) galaxies identified by \citet{alatalo16}. SPOGs combine strong Balmer absorption with emission line ratios consistent with shocks and inconsistent with star-formation. This selection differs from an E+A or K+A selection by allowing for H$\alpha$ emission, even if much of the emission is unlikely to be from star formation. Most SPOGs have star formation histories similar to  traditionally-selected post-starburst galaxies, yet are on average younger and thus may have higher dust obscuration \citep{french18}. There are several theoretical predictions for a higher TDE rate in young starburst/post-starburst galaxies \citep{madigan18, stone18} which would predict an intrinsically high TDE rate in SPOGs as well, although the observed rate might be lower due to the extra dust.

The right panel of Figure~\ref{fig:host_lines} shows a BPT diagram of \host{} and several other TDE hosts along with a comparison sample from the SDSS main spectroscopic survey \citep{strauss02}. The host of ASASSN-19bt lies in the Seyfert or non-star-forming portion of the diagram, along with several other TDE hosts, but it lies outside the region enclosing galaxies with line ratios consistent with shocks. Thus, while it is similar in many ways to SPOGS, it is likely not a member of this population.

We fit the spectral energy distribution (SED) of the host galaxy using the Fitting and Assessment of Synthetic Templates \citep[\textsc{fast};~][]{kriek09} code. We used 
the  {\swift} UVOT UV and $U$, APASS $BVgri$, 2MASS $JHK_S$, and WISE $W1$ and $W2$ magnitudes given in Table~\ref{tab:host_mags} to constrain the model. We assumed a \citet{cardelli89} extinction law with $R_V=3.1$, a Galactic extinction of $A_V = 0.336$ mag \citep{schlafly11}, a Salpeter initial mass function, an exponentially declining star-formation history, and stellar population models from \citet{bruzual03} for the fit. The \textsc{fast} fit indicates that \host{} has a stellar mass of $M_{\star}=1.1^{+1.3}_{-0.1} \times 10^{10}$ M$_{\odot}$, an age of $3.2^{+5.8}_{-0.1}$ Gyr, and a star-formation rate of $\textrm{SFR}=1.7^{+0.6}_{-0.1} \times 10^{-1}$ M$_{\odot}$~yr$^{-1}$. Using the average stellar-mass-to-bulge-mass ratio from the hosts of ASASSN-14ae, ASASSN-14li, and ASASSN-15oi \citep{holoien14b,holoien16a,holoien16b} to scale the stellar mass of the host, as we have done with previous TDEs \citep[e.g.,][]{holoien18b}, results in an estimated bulge mass of $M_B\simeq10^{9.4}$~{\msun}. This corresponds to an estimated black hole mass of $M_{BH}=10^{6.8}$~{\msun} \citep{mcconnell13}, comparable to the BH mass estimates for other TDE hosts \citep[e.g.,][]{holoien14b,holoien16a,holoien16b,brown17a,wevers17,mockler18}.

\subsection{ASAS-SN light curve}
\label{sec:asassn_obs}


\begin{figure*}
\begin{minipage}{\textwidth}
\centering
{\includegraphics[width=0.95\textwidth]{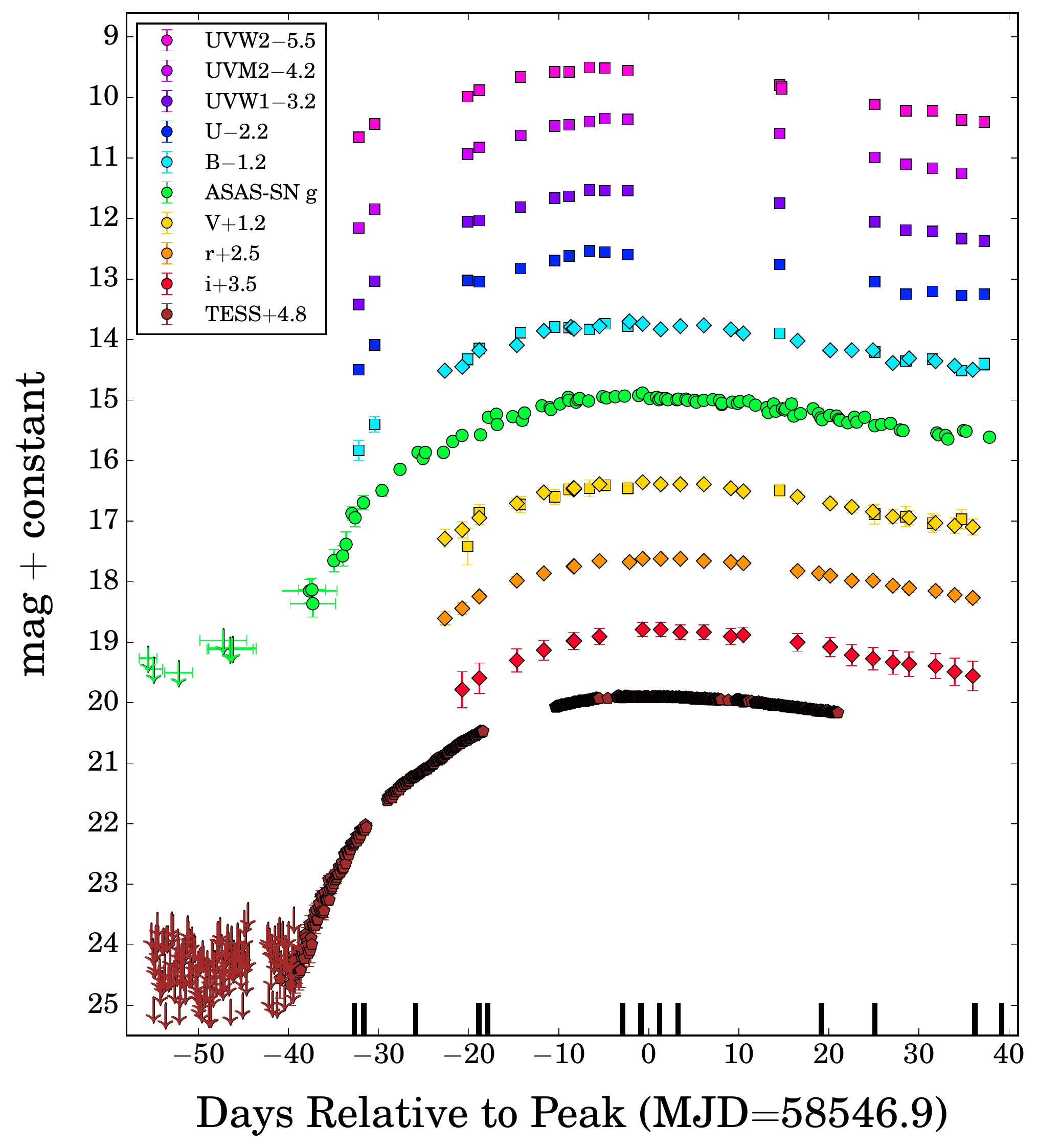}}
\caption{Host-subtracted UV and optical light curves of ASASSN-19bt, showing the ASAS-SN ($g$, circles), {\swift} (UV+$UBV$, squares), Las Cumbres Observatory 1-m telescopes ($BVri$, diamonds) and TESS (pentagons) photometry and spanning from roughly 60 days prior to peak brightness (MJD$=$58546.9) to 35 days after. Arrows indicate $3\sigma$ upper limits for epochs where no transient flux is detected. {\swift} UVOT $B$ and $V$ data were converted to Johnson $B$ and $V$ magnitudes to enable direct comparison with other data. Error bars on the time axis for pre-discovery ASAS-SN data indicate the date range of observations combined to obtain deeper limits and higher signal-to-noise detections. The TESS light curve shows the median magnitude of observations in 2-hour bins, with epochs with negative subtracted flux or prior to our inferred time of first light (MJD$=$58504.6, see Section~\ref{sec:lc_anal}) converted into $3\sigma$ upper limits. Black bars along the X-axis show epochs of spectroscopic follow-up. All data are corrected for Galactic extinction and are presented in the AB system.}
\label{fig:lc}
\end{minipage}
\end{figure*}

ASAS-SN images are processed in real-time using a fully automated pipeline incorporating image subtraction that is performed with the ISIS package \citep{alard98, alard00}. As ASAS-SN monitors the TESS fields with an increased cadence, the field containing ASASSN-19bt was observed on average $1-2$ times per night before and after discovery of the transient with cameras in our ``Cassius'', ``Paczynski'', and ``Payne-Gaposchkin'' units, located in Chile and South Africa. To ensure that no transient flux was present in the reference images used for image subtraction, we constructed a reference image for each camera that observed ASASSN-19bt using only data obtained earlier than 2018 December 1, roughly 60 days prior to discovery. We used these references as templates to subtract the background and host emission from all data taken after 2019 January 1, ensuring that we capture the full rise of the TDE in ASAS-SN data.

We performed aperture photometry on each host-subtracted image using the {\sc IRAF} {\tt apphot} package and an aperture 3 pixels in radius (roughly equivalent to 21\farcs{0}). We calibrated the ASAS-SN $g$-band magnitudes using several stars near the transient with magnitudes available in APASS. For many pre-discovery epochs, when ASASSN-19bt was very faint or not detected, we combined images taken over several days on the same camera to improve the signal-to-noise of our detections and obtain deeper limits on the TDE emission. All ASAS-SN photometry, including both detections and $3\sigma$ upper limits, is presented in Table~\ref{tab:phot}, and we show the ASAS-SN light curve in Figure~\ref{fig:lc} along with our follow-up photometry and photometry from TESS.

While SPOGs typically do have a higher amount of dust obscuration, and the host emission lines indicate that the emission line regions may have a higher amount of obscuration, our host SED fit is sufficiently good to not require additional dust correction. Furthermore, the fact that we clearly detect the TDE in UV filters, and the fact that TDE SED is well-fit by a blackbody without any additional dust correction (see Figure~\ref{fig:sed_evol}) implies that any potential host extinction along the line-of-sight to the TDE must be minimal. For this reason, while we correct all our photometry for Galactic extinction, we do not apply any host extinction correction to our measurements.

\subsection{TESS Observations}
\label{sec:tess_obs}

Located in the TESS CVZ near the South Ecliptic Pole, ASASSN-19bt has been observed by TESS almost constantly since science operations commenced in late July of 2018. The location of the TDE fell within a chip gap during Sector 6 observations, but this is the only Sector in Cycle 1 for which TESS did not observe it. There are 5 full Sectors of pre-disruption observations and a full orbit's worth of observations obtained in Sector 7 prior to first light from the TDE. After first light, the transient's rise is continuously sampled by TESS throughout the remainder of Sector 7 and all of Sector 8, with the exception of a gap in Sector 8 caused by an instrument anomaly on the spacecraft. The transient's epoch of maximum brightness and initial decline are captured in the Sector 9 observations.

As with the ASAS-SN data, we used the ISIS package \citep{alard98,alard00} to perform image subtraction on the TESS full frame images (FFIs) to produce high fidelity light curves. Due to the large pixel scale of the TESS observations, we elected not to rotate a single reference image for use across various sectors and instead chose to construct independent reference images for each sector. This was achieved by selecting the first 100 FFIs of good quality obtained during that sector, excluding those with sky background levels or PSF widths above average for the sector. We entirely exclude any FFIs with data quality flags from our analysis. We adopted additional conservative quality cuts, excluding FFIs obtained when the spacecraft's pointing was compromised, when TESS was either impacted by or recovering from an instrument anomaly, or when significant background effects due to scattered light were present in the images.

\begin{figure*}
\begin{minipage}{\textwidth}
\centering
{\includegraphics[width=0.95\textwidth]{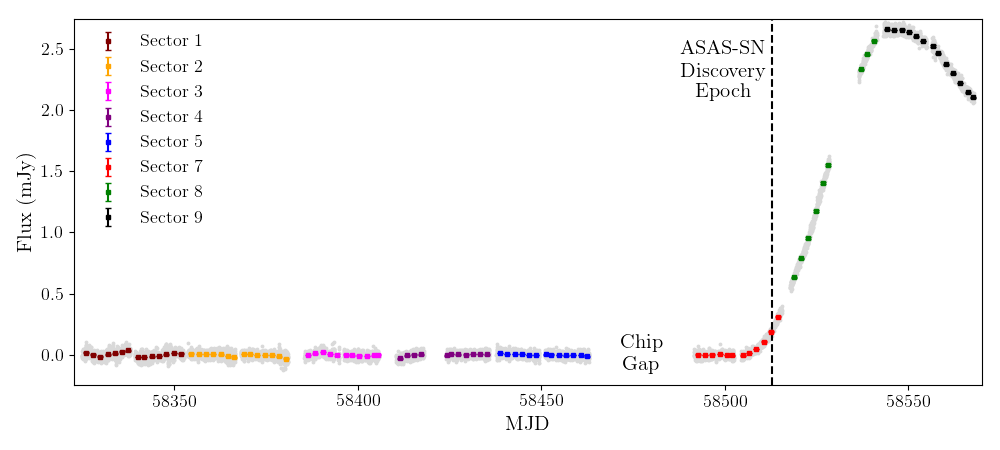}}
\caption{The TESS image subtraction light curve of ASASSN-19bt obtained for all currently available TESS Sectors. ASASSN-19bt was not observed in TESS Sector 6 as the transient coordinates fell on a TESS chip gap. The extended gap in Sector 8 observations is due to an instrument anomaly on the spacecraft. Flux values for every epoch are shown in light gray, and the colored points show the mean values of 2-day bins. Error bars are shown for all of colored points, but are considerably smaller than the symbols.}
\label{fig:TotTESS}
\end{minipage}
\end{figure*}

Because a considerable amount of flux from the TDE is present in the images used to construct the Sector 8 reference, fluxes in the raw difference light curve for this sector are systematically lower than the intrinsic values. We correct for this by applying an offset, which we calculate by matching the first 12 hours of Sector 8 observations (24 epochs) to an extrapolation of the Sector 7 power-law fit shown in Figure~\ref{fig:PowerLawFit} (see Section~\ref{sec:lc_anal}). We subsequently take a simple linear fit between the last 12 hours of Sector 8 observations and the first 12 hours of Sector 9 observations to calculate the offset for the Sector 9 light curve. The measured fluxes were converted into TESS-band magnitudes using an instrumental zero point of 20.44 electrons per second in the FFIs, based on the values provided in the TESS Instrument Handbook \citep{TESSHandbook}. TESS observes in a single broad-band filter, spanning roughly 6000--10000\,\AA{} with an effective wavelength of $\sim$7500 \AA{}, and TESS magnitudes are calibrated to the Vega system \citep{sullivan15}. The full, host-subtracted light curves for all currently available TESS Sectors are shown in Figure~\ref{fig:TotTESS}.


\begin{deluxetable}{cccc}
\tabletypesize{\footnotesize}
\tablecaption{Host-Subtracted Photometry of ASASSN-19bt}
\tablehead{
\colhead{MJD} &
\colhead{Filter} &
\colhead{Magnitude} &
\colhead{Telescope/Observatory} }
\startdata
58491.66 & TESS & >18.84 & TESS \\
58491.74 & TESS & >20.48 & TESS \\
58491.82 & TESS & >19.10 & TESS \\
... & & & \\
58578.42 & $UVW2$ & $15.72\pm0.05$ & \swift \\
58581.62 & $UVW2$ & $15.87\pm0.05$ & \swift \\
58584.15 & $UVW2$ & $15.90\pm0.07$ & \swift \\
\enddata 
\tablecomments{Host-subtracted magnitudes and $3\sigma$ upper limits for all photometric follow-up data. The Telescope/Observatory column indicates the source of the data for each epoch: ``ASAS-SN'' is used for ASAS-SN survey data, ``TESS'' is used for TESS data, ``LCOGT\_1m'' is used for data from the Las Cumbres Observatory 1-m telescopes, and ``\swift'' is used for {\swift} UVOT data. A range of dates given in column 1 indicates the dates of the earliest and latest observations in a set that were combined to improve signal-to-noise. All measurements have been corrected for Galactic extinction and are presented in the AB system. TESS data has been binned in 2-hour bins, as described in Section~\ref{sec:tess_obs}, with the MJD at the center of the bin given in Column 1. Only a portion of this Table is shown here, for guidance regarding its form and content; the entire table is published in machine-readable format in the online journal.}
\label{tab:phot} 
\end{deluxetable}

For comparison with our other photometry, we binned the TESS data for epochs after $\hbox{MJD}=58491$ in 2-hour bins by taking the median flux in each bin and calculating the TESS magnitude from the median flux. We converted the TESS Vega magnitudes to the AB system using an offset of $m_{AB}-m_{Vega} = 0.411$ to match our other photometry. TESS magnitudes and $3\sigma$ limits are presented in Table~\ref{tab:phot} and shown in Figure~\ref{fig:lc}.

\subsection{{\swift} UVOT Observations}
\label{sec:swift}

Following our initial classification of ASASSN-19bt as a possible TDE, we requested and were awarded 17 epochs of target-of-opportunity (TOO) observations with the {\swift} UVOT and XRT, with the first epoch of observations obtained within 2 days of discovery. 

UVOT observations were obtained in the $V$ (5468 \AA), $B$ (4392 \AA), $U$ (3465 \AA), $UVW1$ (2600 \AA), $UVM2$ (2246 \AA), and $UVW2$ (1928 \AA) filters \citep{poole08} in most epochs, with some epochs having only a subset of filters due to scheduling. Each epoch of UVOT data contains 2 observations in each filter, and, in order to measure transient fluxes for each epoch, we first combined the two images in every filter using {\tt uvotimsum}. We then used {\tt uvotsource} to measure counts from the source in the combined images using a 5\farcs{0} aperture and a background region of $\sim$~40\farcs{0} radius to estimate and subtract the sky background. As with the archival data, we then converted the count rates to magnitudes and fluxes using the most recent UVOT calibration \citep{poole08,breeveld10}.


We corrected all the UVOT photometry for Galactic extinction using a \citet{cardelli89} extinction law. To enable the UVOT $B$ and $V$ data to be directly compared with Johnson $B$ and $V$ data obtained from ground-based telescopes, we converted the UVOT $B$- and $V$-band data to Johnson $B$ and $V$ magnitudes using publicly available color corrections\footnote{\url{https://heasarc.gsfc.nasa.gov/docs/heasarc/caldb/swift/docs/uvot/uvot_caldb_coltrans_02b.pdf}}. Finally, we then subtracted the host flux measured from the archival {\swift} observations (UV filters and $U$) or taken from the APASS catalog ($BV$) from each UVOT observation to isolate the transient flux in each epoch. The host-subtracted {\swift} UVOT photometry is presented in Table~\ref{tab:phot} and shown in Figure~\ref{fig:lc}. ASASSN-19bt has one of the brightest measured peak UV magnitudes of any TDE to-date, and is comparable in brightness to ASASSN-14li \citep{holoien16a} despite being more distant. ASASSN-19bt also has 10 epochs of UV observations obtained prior to peak, making this one of the best-sampled multiwavelength rising light curves obtained to-date of a TDE.

\subsection{Other Photometric Observations}
\label{sec:other_obs}

In addition to the ASAS-SN, {\swift}, and TESS observations, we also obtained photometric observations in the $BVri$ filters from Las Cumbres Observatory 1-m telescopes located in Cerro Tololo, Chile, Siding Spring, Australia, and Sutherland, South Africa \citep{brown13}. After applying photometric calibrations, we measured aperture magnitudes using the IRAF {\tt apphot} package, with a 5\farcs{0} aperture region used to extract source counts and a 15\farcs{0}$-$30\farcs{0} annulus used to estimate and subtract background counts. We calibrated the Las Cumbres Observatory magnitudes using several stars in the field with magnitudes available in the APASS DR 10 catalog.

As we did with the UVOT observations, we corrected the ground-based aperture magnitudes for Galactic extinction and subtracted the host flux in each band taken from APASS. We find excellent agreement between the UVOT $B$ and $V$ magnitudes and those measured from the Las Cumbres Observatory telescopes. We present the host-subtracted ground-based photometry in Table~\ref{tab:phot} and show them in Figure~\ref{fig:lc}

\subsection{X-ray Observations}
\label{sec:xray_obs}

\subsubsection{Swift XRT Observations}
\label{sec:swiftxrt}

In addition to the {\swift} UVOT observations, we simultaneously obtained XRT photon-counting (PC) observations of ASASSN-19bt. All observations were reduced following the standard \emph{Swift} XRT data reduction guide\footnote{\url{http://swift.gsfc.nasa.gov/analysis/xrt\_swguide\_v1\_2.pdf}}, with the level one XRT data reprocessed using the \emph{Swift} \textit{xrtpipeline} version 0.13.2 script. Standard filters and screening were applied, along with the most up-to-date calibration files. To increase the signal-to-noise of our observations, we also combined the \textit{Swift} observations into five time bins using \textsc{XSELECT} version 12.9.1c. We used a source region centered on the position of ASASSN-19bt with a radius of 30'', and a source free background region centered at ($\alpha,\delta$) =(06:59:55.2, $-$66:05:37.04) with a radius of 150\farcs{0}. All extracted count rates were corrected for the encircled energy fraction since this source radius contains only $\sim$90\% of the counts from a source at 1.5 keV \citep{moretti04}. 

In most epochs, we do not detect X-ray emission from ASASSN-19bt in the XRT data, and we calculate $3\sigma$ flux limits from the combined XRT data in such cases. However, in one combined set of observations obtained roughly 2 weeks prior to peak light we do detect weak X-ray emission from ASASSN-19bt. We report the X-ray luminosity limits and measured luminosities from the combined {\swift} data in Table~\ref{tab:xray}.

\subsubsection{XMM-Newton Observations}
\label{sec:xmm}
Since the {\swift} XRT observations showed evidence of weak X-ray emission as the source rose to peak, we requested two deep (41.4 ks each) \textit{XMM-Newton Observatory} target of opportunity observation (TOO) of the source. The first observation was taken on 2019 March 1st (ObsID: 0831791001, PI: Auchettl), approximately 3.5 days before peak (MJD$_{\rm XMM_{1}}$=58543.21), while the second observation was taken on 2019 April 15th (ObsID: 0831791101, PI: Auchettl), approximately 42 days after peak (MJD$_{\rm XMM_{2}}$=58589). Both the MOS and PN detectors were used for this analysis and both detectors were operated in full-frame mode using a thin filter. All data reduction and analysis was done using the \textit{XMM-Newton} science system (SAS) version 15.0.02 with the most up to date calibration files.

Due to the fact that XMM-Newton suffers from periods of high background and/or proton flares that may affect the quality of the data, we checked for these periods by generating a count rate histogram of the events that have energies between 10$-$12 keV. We find that our observations are only minimally affected by background flares, giving an effective exposure in the PN and MOS detectors of 32 ks and 37 ks for the first observation and 40 ks and 39 ks for the second observation.

For our analysis we used the standard screening of events, with single to quadruple events (PATTERN $\leq$ 12) chosen for the MOS detectors. For the PN detector, only single and double events (PATTERN$\leq$4) were selected. We also used the standard screening FLAGS for both the MOS (\#XMMEA EM) and PN (\#XMMEA EP) detectors. We corrected for vignetting by processing all event files using the task \textsc{evigweight}. We extracted spectra of ASASSN-19bt from both the MOS and PN detectors using the SAS task \textsc{evselect} and the cleaned event files from all detectors. We used the same source region that was used to analyze the \textit{Swift} observations. To avoid chip gaps we used a smaller background region centered at ($\alpha,\delta$) =(7:00:38.535, $-$66:05:43.21) with a radius of 72\farcs{0} for the first observation and ($\alpha,\delta$) =(7:00:06.415, $-$66:06:31.27) with a radius of 72\farcs{0} for the second observation. To increase the signal-to-noise of the MOS spectra, we combined these spectra together using the SAS task \textsc{epicspeccombine}. To analyze the spectra extracted from our \textit{XMM-Newton} observations, we used the X-ray spectral fitting package (XSPEC) version 12.10.1f and $\chi^2$ statistics. Using the \textsc{FTOOLS} command \textit{GRPPHA}, we grouped both the PN and merged MOS spectra with a minimum of 10 counts per energy bin. These data are further discussed in Section \ref{sec:xray_anal}, and the X-ray luminosity measured from the \textit{XMM-Newton} observations is given in Table~\ref{tab:xray}.



\subsection{Spectroscopic Observations}
\label{sec:spec_obs}

After obtaining our first classification spectrum of ASASSN-19bt, we began a program of spectroscopic follow-up to complement our photometric dataset. Our follow-up spectra were obtained with LDSS-3 on the 6.5-m Magellan Clay telescope, the Inamori-Magellan Areal Camera and Spectrograph \citep[IMACS;][]{dressler11} on the 6.5-m Magellan-Baade telescope, the Goodman Spectrograph \citep{clemens04} on the Southern Astrophysical Research (SOAR) 4.1-m telescope, and the Wide Field Reimaging CCD Camera (WFCCD) mounted on the du Pont 100-inch telescope. These observations included 7 spectra obtained prior to peak light and 4 spectra obtained within 3 days of peak light.

The majority of our spectra were reduced and calibrated using standard procedures in \textsc{Iraf}, including bias subtraction, flat-fielding, 1-D spectroscopic extraction, and wavlength calibration via comparison to an arc lamp. The IMACS data from 2019 February 07 were reduced using an updated version of the routines developed by \citet{kelson14} using He and Hg lamps for wavelength calibration. We flux calibrated our observations using standard star spectra obtained on the same nights as the science spectra and masked prominent telluric features. Details of the spectra in our dataset are presented in Table~\ref{tab:spec_details}.


\begin{figure*}
\begin{minipage}{\textwidth}
\centering
{\includegraphics[width=0.90\textwidth]{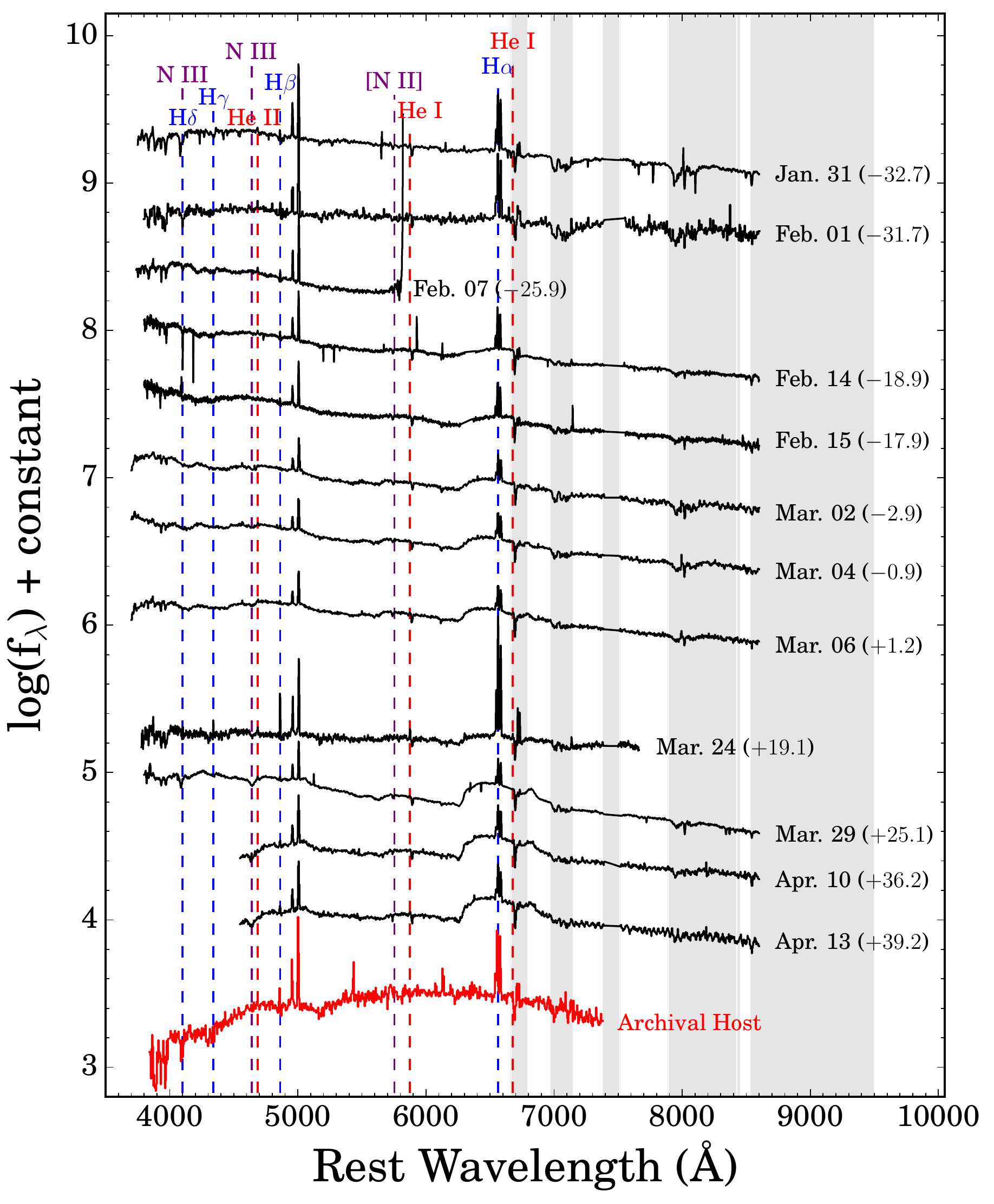}}
\caption{Spectroscopic evolution of ASASSN-19bt spanning from 33 days prior to peak (2019 March 04) through 36 days after peak. All spectra have been flux calibrated using our photometry, as described in Section~\ref{sec:spec_obs}. The date for each spectrum and phase relative to peak in observed days is shown next to each spectrum. Hydrogen, helium, and nitrogen features that are common to TDEs are indicated with blue, red, and purple dashed lines, respectively. Telluric bands are shown in light gray. The calibrated archival host spectrum from 6dF is shown in red.}
\label{fig:spec_evol}
\end{minipage}
\end{figure*}

We further calibrated our spectra using our photometric dataset. To obtain magnitudes with a similar amount of host contamination as the spectra, which were observed through slits of roughly 1\farcs{0} width, we measured small aperture magnitudes from our {\swift} and Las Cumbres Observatory data, using a 1\farcs{5} aperture for the Las Cumbres Observatory data and a 3\farcs{5} aperture for the {\swift} data, due to the larger pixel scale of {\swift}. For each filter that was completely contained in the wavelength range covered by a given spectrum and for which we could either interpolate the small aperture light curves or extrapolate them by 1 hour or less we extracted synthetic photometric magnitudes from the spectrum. As the {\swift} data has a larger PSF and the small aperture magnitudes are more uncertain, we calibrated the spectra using only Las Cumbres Observatory data, except for the three spectra taken prior to 2019 February 10, when we obtained our first observations with Las Cumbres Observatory telescopes. We then fit a line to the difference between the synthetic and observed fluxes as a function of central filter wavelength and scaled each spectrum by the photometric fits. Finally, we corrected each spectrum for Galactic reddening using a Milky Way extinction curve assuming $R_V=3.1$ and $A_V=0.336$ \citep{schlafly11}. We also used the procedure to calibrate the archival 6dF spectrum to the same flux scale as our follow-up spectra using the archival magnitudes shown in Table~\ref{tab:host_mags} for calibration. 

Figure~\ref{fig:spec_evol} shows the spectroscopic evolution of ASASSN-19bt as well as the calibrated host spectrum. Prominent telluric bands have been marked in the Figure, with the feature from 7550\AA$-$7720\AA\ and chip gaps (where present) masked to facilitate plotting. Similar to what was seen with PS18kh \citep{holoien18a}, there is little evidence of broad lines prior to peak light, with the lines becoming more prominent after peak. We further analyze the line emission in Section~\ref{sec:spec_anal}.


\section{Analysis}
\label{sec:analysis}

\subsection{Position, Redshift, and $t_{Peak}$ Measurements}
\label{sec:params}

We measured the position of ASASSN-19bt using our initial $V$-band image and a $V$-band image taken near peak from the Las Cumbres Observatory 1-m telescopes. Using the early image as a subtraction template, we generated a subtracted image of the TDE. While some TDE flux was likely removed along with the host flux, this allowed us to measure a centroid position of only the TDE signal. Using the \textsc{Iraf} task \texttt{imcentroid} we measured the centroid position of the flux in the subtracted image as well as the centroid position of the nucleus of the host galaxy in the early $V$-band image that was used as the subtraction template, which is likely host-dominated. From this method, we obtain a position for ASASSN-19bt of RA$=$07:00:11.41, Dec$=-$66:02:25.16. This is offset by $0\farcs14\pm0\farcs15$ from the position of the host nucleus measured from the early image, which corresponds to a physical offset of $78.2\pm83.8$~pc.

The redshift of the host galaxy in the 6dF catalog is reported as $z=0.0262$. In order to verify this, we downloaded the 6dF spectrum and measured the redshift using the narrow H$\alpha$ and [\ion{O}{3}] 5007/4959\AA{} emission lines, finding $z=0.026$. As this is consistent with the reported redshift, we adopt $z=0.0262$ and the corresponding luminosity distance of $d=115.2$~Mpc throughout our analysis. 

To obtain an estimate of the time of peak light, we fit a parabola to the host-subtracted ASAS-SN $g$-band light curve prior to MJD$=$58560, as the decline of the light curve is flatter than the rise, making a parabolic fit to the entire light curve inaccurate. To estimate the uncertainty on the peak date, we generated 10000 realizations of the $g$ light curve prior to MJD$=$58560 with each magnitude perturbed by its uncertainty assuming Gaussian errors. We then fit a parabola to each of these light curves and calculated the 68\% confidence interval and median $t_{peak}$ values from these realizations. Using this procedure, we find $t_{g,peak}=58546.9\pm0.2$ and $m_{g,peak}=14.9$. We also performed the same analysis for all of our photometric filters and found that there is some evidence that the bluer filters peaked earlier, with $t_{UVW2,peak}=58544.2^{+2.6}_{-1.9}$ and $t_{i,peak}=58548.9^{+2.4}_{-1.5}$, not unlike what has been seen in some other TDEs \citep[e.g.,][]{holoien18a}. Due to the high cadence of the ASAS-SN light curve, its peak is much better constrained than those of the other filters, and we adopt the $g$-band peak of $t_{g,peak}=58546.9$, corresponding to 2019 March 04.9, throughout this paper.

\subsection{Light Curve Analysis}
\label{sec:lc_anal}

\begin{figure}
\includegraphics[width=\columnwidth]{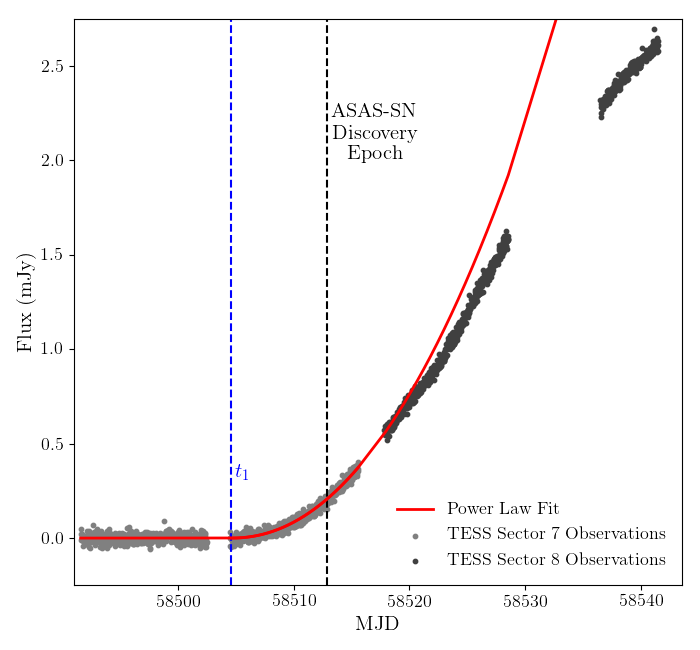}
\caption{The TESS image subtraction light curve of ASASSN-19bt obtained during Sector 7 and Sector 8. Individual flux measurements obtained for each FFI are shown in gray, with different shades denoting each Sector, and a best-fit power law model to the Sector 7 data is shown in red. This power law fit yields a time of first light of $t_1=58504.61\pm0.42$ and has an index of $\alpha=2.10\pm0.12$. The blue dashed line and black dashed line show the inferred $t_1$ and the ASAS-SN discovery date, respectively. The light curve begins to diverge from the initial power law evolution roughly 15 days after first light.}
\label{fig:PowerLawFit}
\end{figure}

We characterized the early-time rise of ASASSN-19bt with a power-law
\begin{flalign}
&f= z \textnormal{ when } t<t_1, \textnormal{ and}&\\
&f= z + h\left( \frac{ t-t_1}{ \hbox{days}}\right)^{\alpha} \textnormal{ when } t>t_1.&
\end{flalign}
model for the TESS flux, described by a residual background $z$, the start of the rise $t_1$, a flux scale $h$ and the power law index $\alpha$. We use the \textsc{scipy.optimize.curve\_fit} package's Trust Region Reflective method to obtain a best fit model with parameters $z=-0.22\pm0.22$ $\mu$Jy, $h=2.44\pm0.90$ $\mu$Jy, $t_1=\textnormal{ MJD}=58504.61\pm0.42$, and $\alpha = 2.10\pm0.12$. This fit is shown as the red curve in Figure~\ref{fig:PowerLawFit}.

This power-law index is consistent with the ``fireball'' model used to model the early flux from SNe where
\begin{equation}
    L_\nu \propto r^2 T \propto v^2 (t-t_1)^2 T,
      \label{eqn:fireball}
\end{equation}
which assumes a band on the Rayleigh-Jeans tail of the SED and homologous expansion. For the early phases of SNe, this power-law rise with $L_\nu \propto (t-t_1)^2$ is due to the fact that the velocity $v$ and temperature $T$ are nearly constant \citep[e.g.,][]{riess99,nugent11}. However, as we shall see in Section~\ref{sec:sed_anal}, the temperature of ASASSN-19bt does not appear to be constant in these early phases, so the consistency of the exponent $\alpha$  with the fireball model appears coincidental.  

Given the distance to the source and the effective wavelength of the TESS band pass, the flux scale given by the fit implies a radius of
\begin{equation}
      r = 10^{13.9} T_4^{-1/2} \left(t-t_1\right)^{1.05}~\hbox{cm}
\end{equation}
for a temperature of $T= 10^4 T_4$~K, the central values of $h$ and $\alpha$ and assuming Eqn.~\ref{eqn:fireball}. For our estimated black hole mass, this corresponds to $100 T_4^{-1/2}$ gravitational radii at $t-t_1=1$~day.  Since the emission region is probably not spherical, it is probably more accurate to say that the surface area producing the UV/optical emission is $4 \pi r^2$.  

The TESS light curve begins to deviate from the initial power-law rise approximately 15 days after first light, with the rise slowing as it approaches peak brightness. Based on the inferred time of first light taken from the power-law fit and the time of peak light measured from the ASAS-SN light curve, we constrain the rise time to $41.2\pm0.5$~rest-frame days.


The photometric precision of the TESS light curve, particularly with the addition of the Sector 9 data, allows us to see clearly that the TDE light curve is very smooth, with little-to-no short-term variability. This is in contrast to what is often seen during AGN flares \citep[e.g.,][]{peterson93,peterson04,shappee14}. While the light curves from ASAS-SN and Las Cumbres Observatory are also quite smooth, the TESS cadence provides us with a unique ability to see exactly how smooth the light curve is. To take advantage of this, we fit the TESS Sector 9 light curve taken after peak light with a power-law decline of the form
\begin{equation}
      f= z - h\left( \frac{ t-t_{peak}}{ \hbox{days}}\right)^{\alpha}
\end{equation}
with $t_{peak}$ being the peak date inferred from the ASAS-SN light curve.


\begin{figure}
\includegraphics[width=\columnwidth]{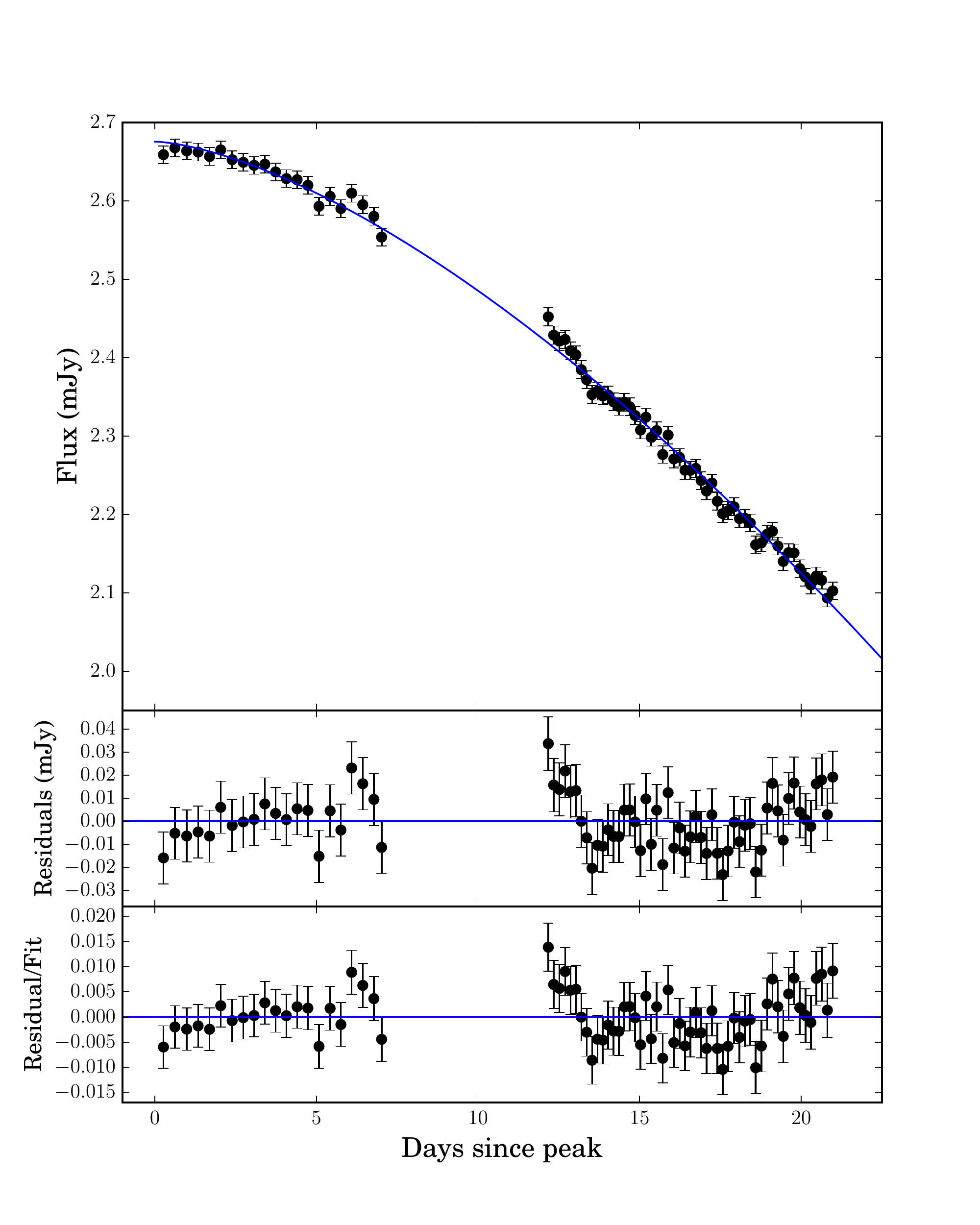}
\caption{The post-peak TESS light curve and best-fit power law model (top panel). The middle and lower panels show the fit residuals and residuals/flux, respectively.}
\label{fig:decline_fit}
\end{figure}

The best-fit power law has the parameters $z=2.68$ $\mu$Jy, $h=0.006$ $\mu$Jy, and $\alpha = 1.53$, with a reduced chi-squared value of $\chi^2_\nu=1.05$. The TESS declining light curve and best-fit model are shown in Figure~\ref{fig:decline_fit}. The fit residuals are roughly 1\% or less of the actual flux, with an RMS value of $0.01$, implying that deviations from the power-law fit are likely systematic. The fact that the light curve is very well-fit by a simple model such as this demonstrates the smoothness of the light curve in a way that was not previously possible.


\begin{figure*}
\begin{minipage}{\textwidth}
\centering
\subfloat{{\includegraphics[width=0.48\textwidth]{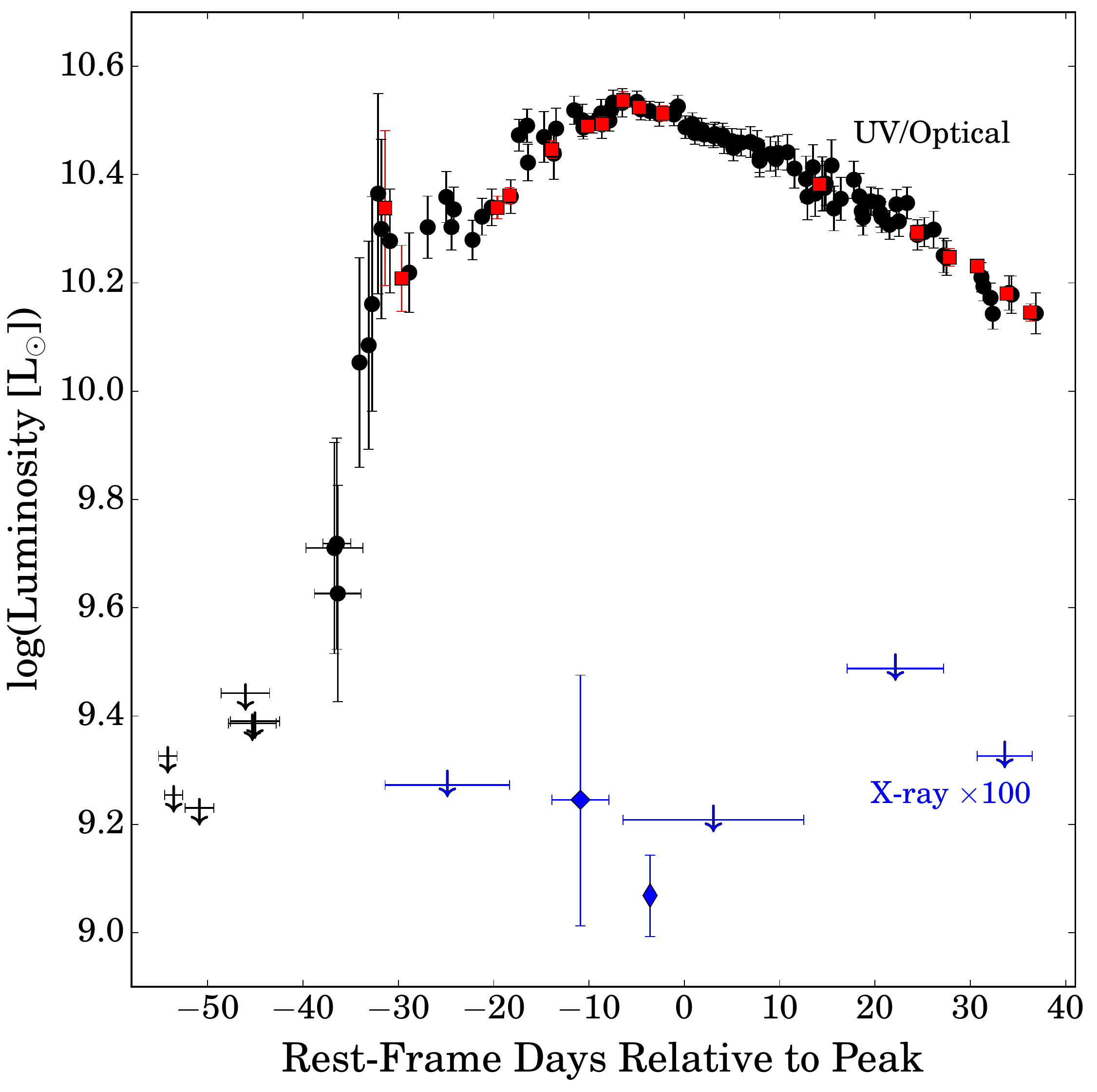}}}
\subfloat{{\includegraphics[width=0.48\textwidth]{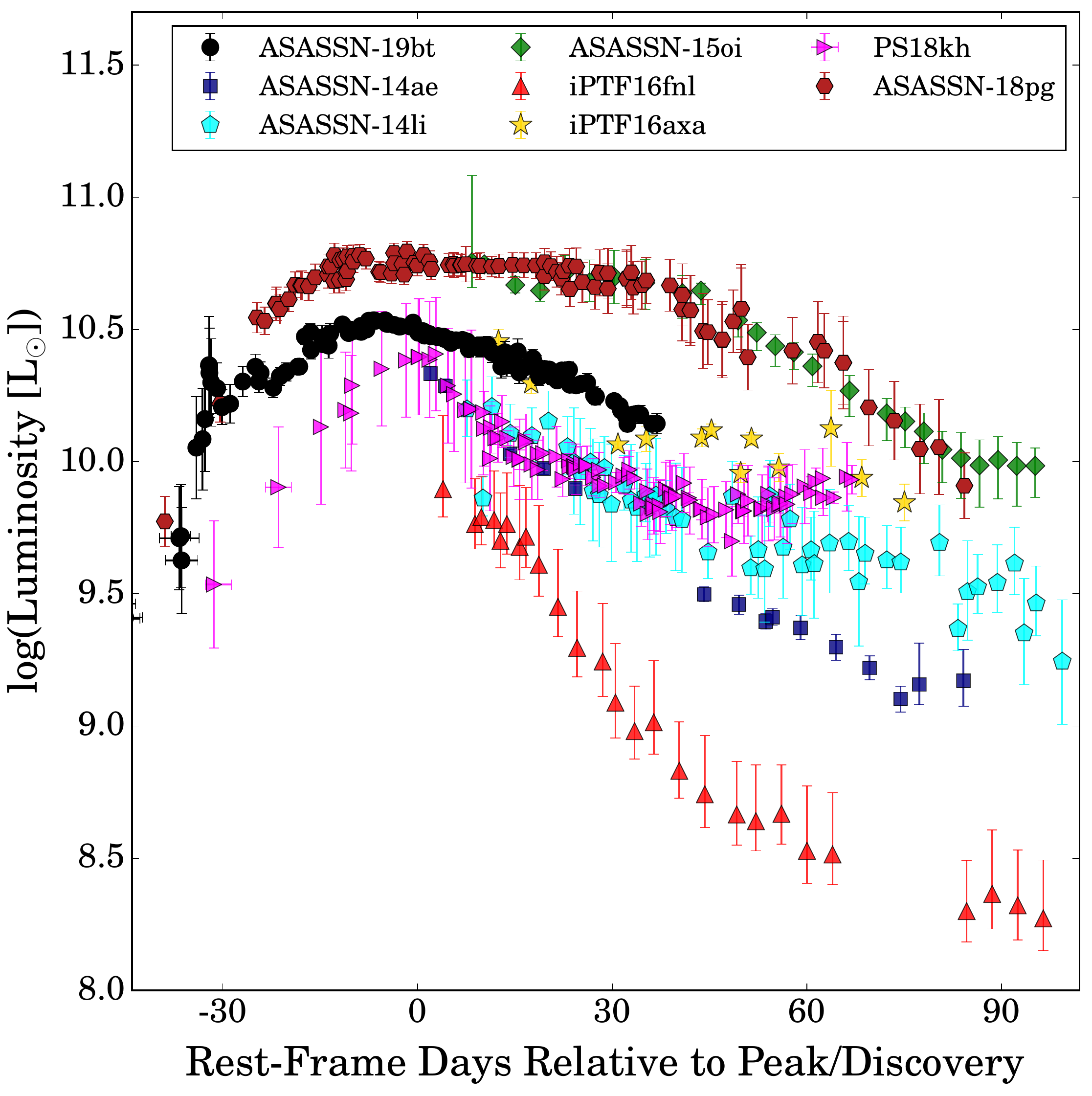}}}
\caption{\emph{Left Panel}: Evolution of the luminosity of ASASSN-19bt from blackbody fits to the UV/optical {\swift} SED (red squares) and estimated from the ASAS-SN $g$-band data by applying bolometric corrections based on the {\swift} fits (black circles). Swift XRT and XMM-Newton X-ray luminosities are shown as wide and thin blue diamonds, respectively, multiplied by a factor of 100 to improve readability. X-axis error bars indicate data ranges for data combined to obtain a single measurement, and downward arrows indicate upper limits. {\emph{Right Panel}: Comparison of the luminosity evolution of ASASSN-19bt (black circles) to that of the TDEs ASASSN-14ae \citep[navy squares;][]{holoien14b}, ASASSN-14li \citep[cyan penatgons;][]{holoien16a}, ASASSN-15oi \citep[green diamonds;][]{holoien16b}, iPTF16fnl \citep[red triangles;][]{brown17b}, iPTF16axa \citep[gold stars;][]{hung17}, PS18kh \citep[magenta right-facing triangles;][]{holoien18a}, and ASASSN-18pg (brown hexagons; Holoien et al., in prep.). Time is shown in rest-frame days relative to peak for those objects which have observations spanning the peak of the light curve (ASASSN-19bt, ASASSN-18pg, and PS18kh) and in days relative to discovery for those objects which do not (ASASSN-14ae, ASASSN-14li, ASASSN-15oi, and iPTF16fnl).}}
\label{fig:lum_evol}
\end{minipage}
\end{figure*}

\subsection{SED Evolution}
\label{sec:sed_anal}

To better understand the physical parameters of the transient, we modeled the UV and optical SED of ASASSN-19bt for epochs where {\swift} data were available as a blackbody. We used Markov Chain Monte Carlo methods to fit the blackbody SED, using a flat prior of $10000$~K~$\leq T \leq55000$~K so as not to overly influence the fits. From the blackbody fits, we estimate the bolometric luminosity, temperature, and radius of ASASSN-19bt in each epoch.

To take better advantage of the high-cadence light curve from ASAS-SN, we used the {\swift} blackbody fits to calculate bolometric corrections for $g$-band magnitudes by linearly interpolating between the previous and next $g$-band measurements bracketing each epoch of {\swift} observations. We then estimated the bolometric luminosity of ASASSN-19bt from the ASAS-SN light curve, by linearly interpolating bolometric corrections calculated for the {\swift} epochs to each epoch of $g$ data. For times prior to our first {\swift} observation, we use the bolometric correction from the first {\swift} SED fit. The luminosity evolution calculated from the {\swift} SED fits and estimated from the $g$-band light curve is shown in the left panel of Figure~\ref{fig:lum_evol}.

The luminosity fits and corrected $g$-band light curve indicate that both the rise to peak and then initial decline after peak are relatively smooth. However, the first {\swift} SED fit indicates that the luminosity was higher in the first epoch than it was in the second, resulting in a short decline at $t\simeq-32$~rest-frame days before peak. This corresponds to a similar drop in the early temperature (see below). Such behavior has not been seen in TDEs prior to ASASSN-19bt. However, {\swift} UV data has not been obtained at such early times prior to peak for previous TDEs. This makes it unclear whether an early, rapid drop in temperature and luminosity is common in TDEs, or unique to ASASSN-19bt, and further highlights the need for early detection and prompt scheduling of follow-up observations. 

The early luminosity spike is not seen in the un-corrected $g$-band data, and the rapid rise to this early peak shown in Figure~\ref{fig:lum_evol} is driven by the fact that we used the first {\swift} luminosity fit to calculate the bolometric correction for all the previously obtained epochs of $g$-band data. Due to the 10-day gap between our second and third {\swift} observations, it is also difficult to determine precisely when the luminosity begins to rise again. However, we can say that the early luminosity spike must have lasted for at least 2 days, and could have been as long as roughly 15 days, depending on how rapid the rise actually was and when it stopped declining and began to re-brighten. We also note that while the luminosity began to re-brighten sometime between $-30$ and $-20$ days prior to peak, its temperature continued to drop during this time, resulting in an early temperature decline that lasted at least 12 rest-frame days.

In the right panel of Figure~\ref{fig:lum_evol}, we show a comparison of the light curve evolution of ASASSN-19bt to that of several other TDEs from the literature: ASASSN-14ae \citep{holoien14b}, ASASSN-14li \citep{holoien16a}, ASASSN-15oi \citep{holoien16b}, iPTF16fnl \citep{brown17b}, iPTF16axa \citep{hung17}, PS18kh \citep{holoien18a}, and ASASSN-18pg \citep[][ Holoien et al. (in prep.)]{leloudas19}. The luminosity rise of ASASSN-19bt is similar to those of ASASSN-18pg and PS18kh with the exception of the spike at $t=-32$ days, and it begins to decline more rapidly after peak than ASASSN-18pg, which has a pleateau-like phase at peak. Both the decline rate shortly after peak and the total luminosity are similar to most of the other objects in the sample, and ASASSN-19bt peaked at a luminosity of $L\simeq1.3\times10^{44}$~ergs~s$^{-1}$.

Figure~\ref{fig:lum_evol} also shows the X-ray luminosities calculated from the {\swift} XRT and XMM-Newton observations, scaled up by a factor of 100 to make them comparable to the UV/Optical luminosities. We do not detect X-ray emission in most epochs, and all detected luminosities are 3 or more orders of magnitude weaker than the UV/optical emission. We also see some evidence that X-rays are only detected near the peak of the UV/optical light curve. We further analyze and discuss the X-ray results in Section~\ref{sec:xray_anal} below.

\begin{figure}
\centering
\includegraphics[width=0.45\textwidth]{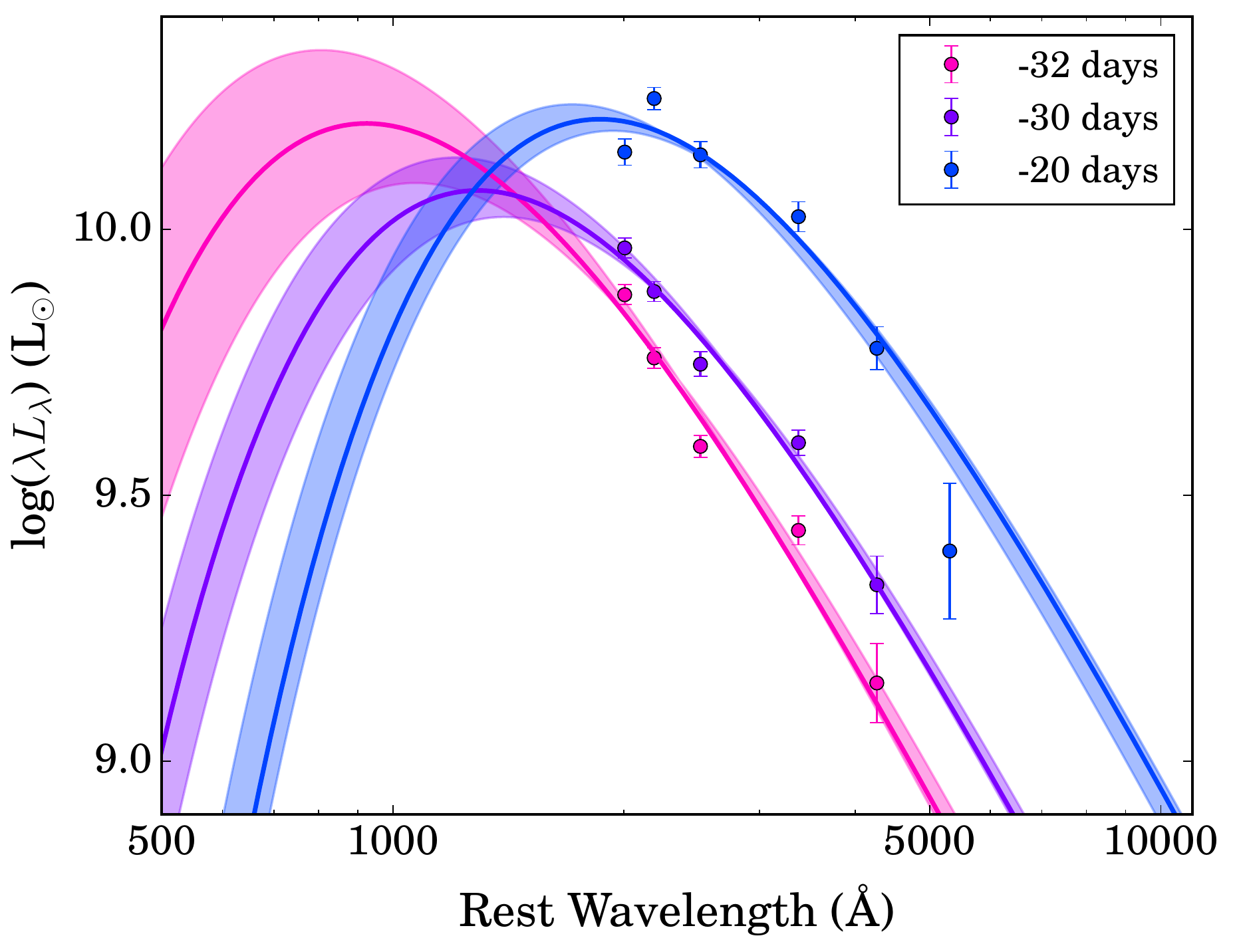}
\caption{The first three {\swift} epochs and their blackbody fits. The solid lines show the SEDs of the median luminosity fits to each epoch, and the shaded regions show the range of SEDs for the $16-84$\% confidence interval on the luminosity. The phase in rest-frame days relative to peak is given in the legend.}
\label{fig:sed_evol}
\end{figure}

We show the SEDs and blackbody fits for the first three {\swift} epochs in Figure~\ref{fig:sed_evol}. The SED for the median luminosity and the range corresponding to the $16-84$\% confidence interval on the luminosity are also shown. The emission clearly becomes redder over these 12 days, with the $UVW2$ flux becoming fainter relative to the $UVM2$ flux over time. After $-20$ days, the TDE continues to exhibit the highest luminosity in the $UVM2$ filter, which drives the cooler temperature fits up to and shortly after peak.

Integrating the entire rest-frame bolometric light curve, including both the {\swift} blackbody fits and the converted $g$-band data, we obtain a total radiated energy of $E=(5.92\pm0.06)\times10^{50}$~ergs. Of this, $(3.17\pm0.05)\times10^{50}$~ergs are released during the rise to peak, indicating that a large fraction of the energy radiated by TDEs can be emitted prior to peak light. The accreted mass required to generate the emitted energy is $M_{Acc}\simeq0.003\eta_{0.1}^{-1}$~\msun, where the accretion efficiency is $\eta=0.1\eta_{0.1}$. This low implied accreted mass is similar to what has been seen in other TDE, and again indicates that either only a small fraction of the bound stellar material is actually accreting onto the SMBH, or that the material accretes with very low radiative efficiency.


\begin{figure}
\centering
\includegraphics[width=0.425\textwidth]{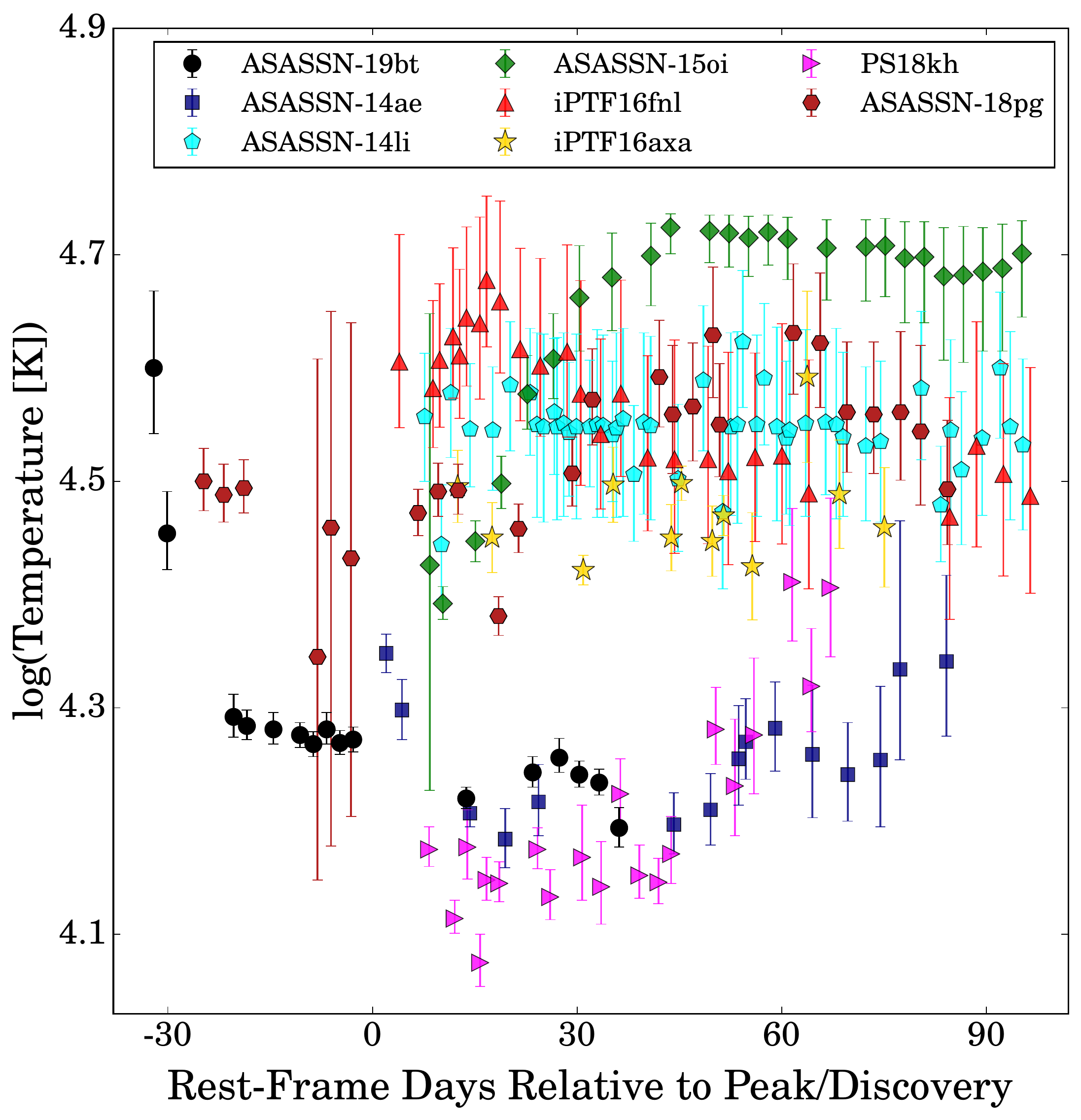}
\caption{Temperature evolution of ASASSN-19bt from the {\swift} blackbody fits (black circles) compared with that of the objects in our TDE comparison sample. Time is in days relative to peak light or relative to discovery, as outlined in the caption of Figure~\ref{fig:lum_evol}, and the symbols and colors match those of Figure~\ref{fig:lum_evol}.}
\label{fig:temp_evol}
\end{figure}

The blackbody temperature evolution of ASASSN-19bt is shown in Figure~\ref{fig:temp_evol}, along with that of the same TDE comparison sample shown in Figure~\ref{fig:lum_evol}. Unique among the TDEs shown in the Figure, ASASSN-19bt exhibits a steep temperature decline in the first 3 epochs, corresponding to the luminosity decline over the same period, before leveling off and exhibiting the relatively constant temperature evolution that is expected. None of the comparison objects have UV data as early as ASASSN-19bt, so it is possible that an early drop in temperature is common, but not observed due to discovering the TDE flare too late. After the initial decline, the temperature remains fairly steady at $T\simeq16000-17000$~K, on the low end of the temperature range observed for TDEs.

\begin{figure}
\centering
\includegraphics[width=0.425\textwidth]{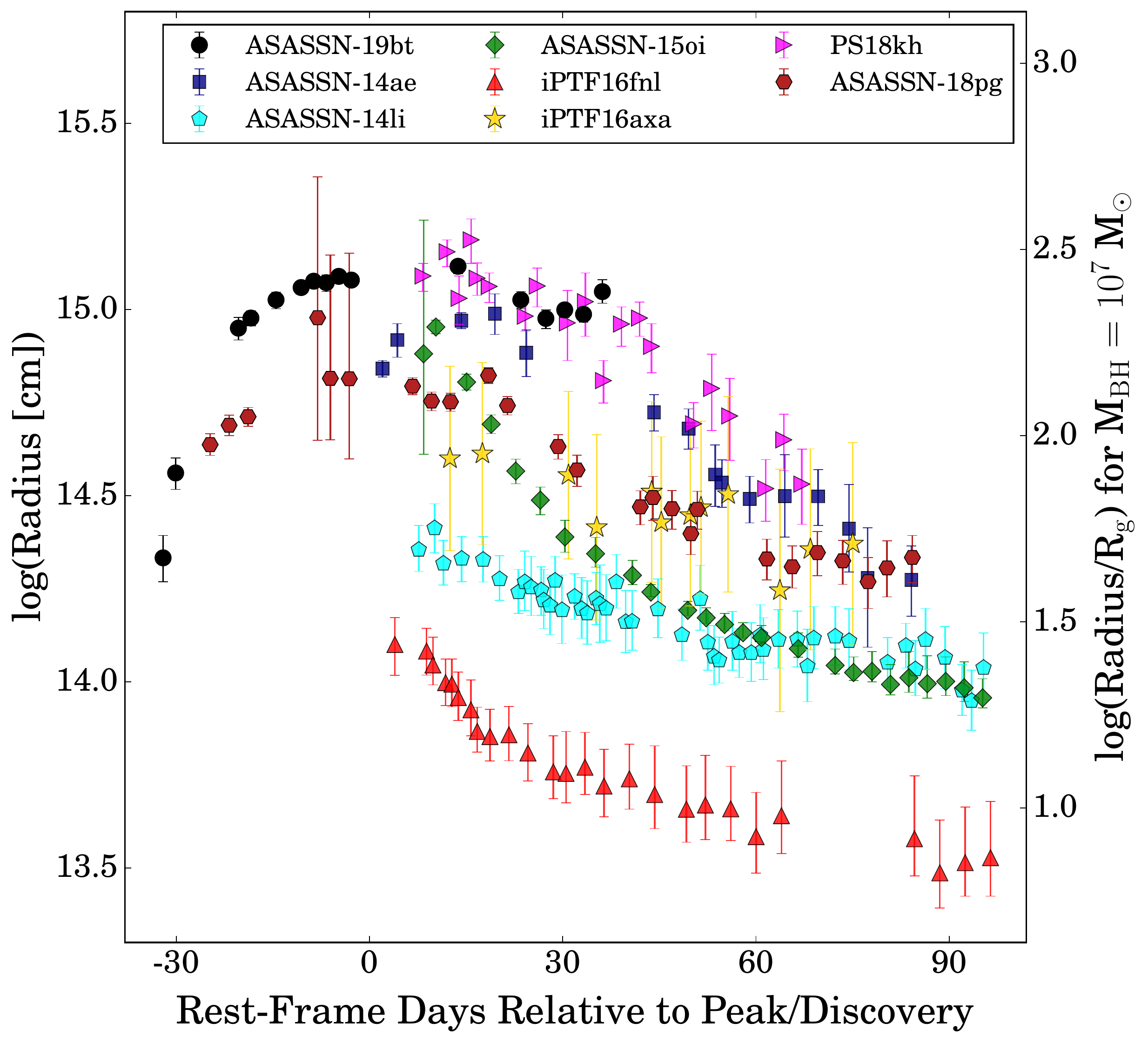}
\caption{Blackbody radius evolution of ASASSN-19bt from the {\swift} fits (black circles) compared with the radius evolution of the objects in our TDE comparison sample. Time is in days relative to peak light or relative to discovery, as outlined in the caption of Figure~\ref{fig:lum_evol}, and the symbols and colors match those of Figure~\ref{fig:lum_evol}. The left-hand scale shows the radius in units of cm, while the right-hand scale shows the same scale in units of the gravitational radius for a $10^7$~\msun~black hole.}
\label{fig:rad_evol}
\end{figure}

Figure~\ref{fig:rad_evol} shows the blackbody radius evolution of ASASSN-19bt taken from the {\swift} fits compared to the radius evolution of the other TDEs in our comparison sample. ASASSN-19bt exhibits a rapidly growing radius in the first 3 epochs to match the drop in temperature over the same timeframe. It then appears to peak and hold relatively steady for the remainder of our period of observation at one of the largest radii of the TDEs in our sample. While there does not seem to be a strong correlation between size and rate of decline in radius, it appears that the hotter TDEs tend to have smaller emitting regions, which is not unexpected given that they are have similar luminosities. ASASSN-19bt is a very close match to PS18kh in both temperature and radius evolution, and it will be interesting to see if ASASSN-19bt exhibits a similar ``re-brightening/plateau'' phase later in its evolution.


\begin{figure}
\centering
\includegraphics[width=0.425\textwidth]{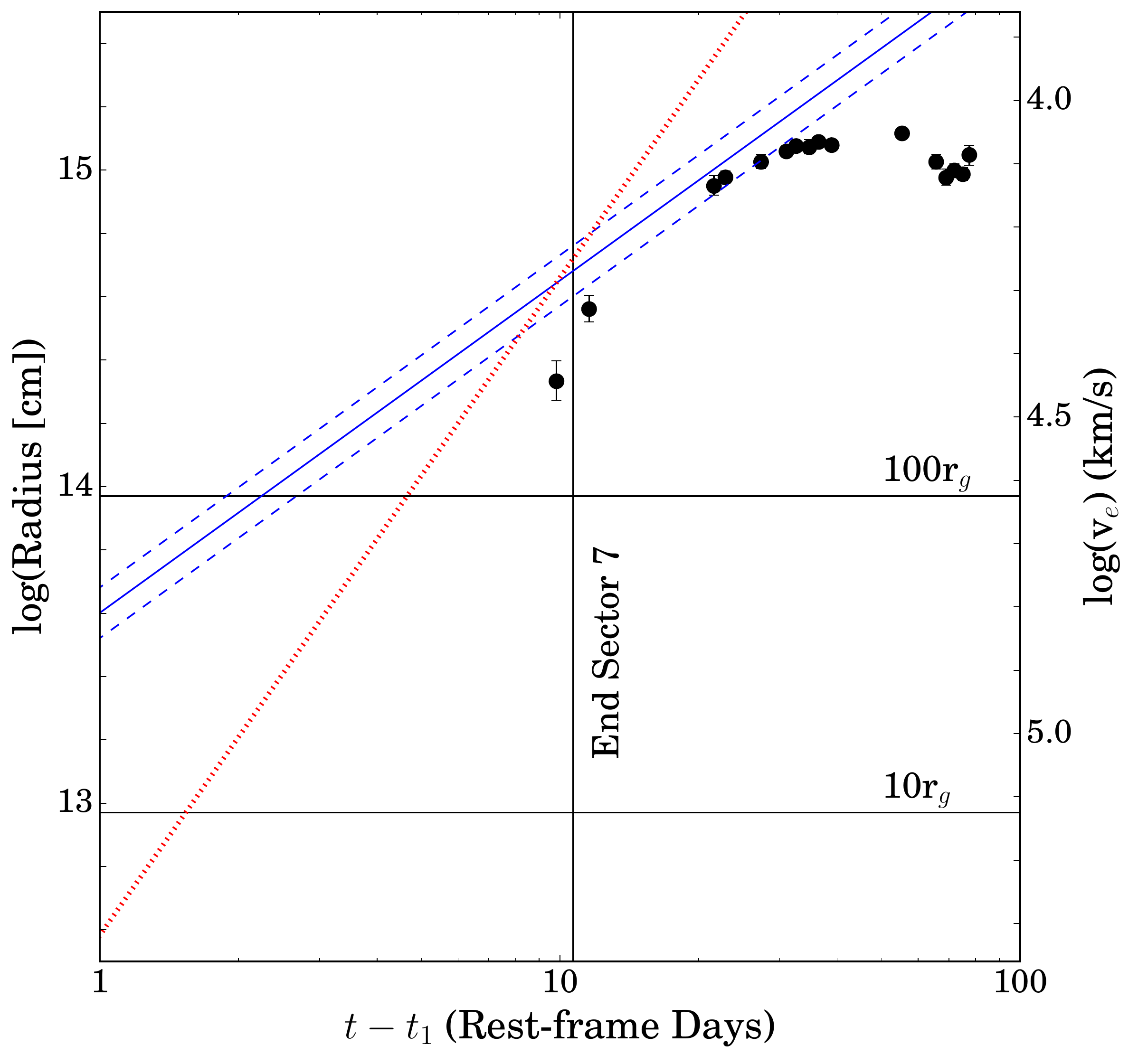}
\caption{Radius evolution in rest-frame days from the estimated start of the TESS emission. The points show the results from the SED fits. The solid blue line bracketed by blue dashed lines shows the radius estimate from the TESS light curve assuming the temperature of the first Swift epoch, while the dotted red line shows the radius estimate assuming a power law decline in temperature based on the first two Swift epochs. The dashed lines illustrate the scale of the uncertainties from the $h$ parameter of the TESS light curve fits. The scale on the right is the escape velocity assuming a BH mass of $10^{6.8}M_\odot$. A vertical line marks the end of TESS Sector 7 and horizontal lines show the radii corresponding to 10 and 100 gravitational radii given our BH mass estimate.}
\label{fig:rad_evol2}
\end{figure}

We can also combine the radius estimates from the SED fits with the results from fitting the TESS light curve in \S\ref{sec:lc_anal}, as shown in Fig.~\ref{fig:rad_evol2}.  Here we work in terms $t-t_1$ based on the onset time found in the TESS fits. The TESS radius estimate depends on the temperature, and we show two simple assumptions. We either assume the temperature found for the first Swift epoch ($\log(T/K)=4.60$) or the power-law in temperature defined by the first two Swift epochs ($\log(T/K)=6.66-2.06\log(t-t_1)$) in order illustrate the uncertainties. Where they overlap, the two sets of radius estimates agree reasonably well (factor of $\sim 2$). At the time of our earliest TESS detection, when $t-t_1\sim 2$~days, the emission region probably had an effective size of some tens of gravitational radii.

Fig.~\ref{fig:rad_evol2} also shows the escape velocity corresponding to a given radius assuming a BH mass of $10^{6.8}M_\odot$. For the constant temperature case, the TESS radius evolution corresponds to a velocity of $\sim 2700$~km/s which is well below the scale of the escape velocity and well above the sound speed implied by the temperatures. With the rapidly evolving temperature, the velocity implied by the TESS radius is $870 (t-t_1)^{1.08}$~km/s, so accelerating from 870~km/s on day $1$ to 10500~km/s on day $10$. Similarly, the $\sim 10^{14.5}$~cm sizes implied by the first few Swift epochs also only require velocities of $3000$-$4000$~km/s.  Thus, if the early time radii implied either by the TESS flux evolution or the first few SED fits are related to the distance from the black hole, the apparent photosphere is expanding very slowly compared to the local escape speed.  If these velocities are related to dynamical velocities, then the emission region must be at a distance near $10^{16.3}$~cm from the BH and the emission region is very small compared to the distance.


\subsection{X-Ray Analysis}
\label{sec:xray_anal}

During the first $\sim$ two weeks of its evolution, our \textit{Swift} XRT observations of ASASSN-19bt showed no evidence of X-ray emission. We first detect the source approximately two weeks prior to the peak, the first such detection.  For those events which do show evidence of X-ray emission (e.g., ASASSN-14li, \citealt{holoien16a,brown17a}; ASASSN-15oi, \citealt{holoien16b,gezari17, holoien18a};  ASASSN-18ul, \citealt{wevers19}; and those in \citealt{auchettl17}), the X-ray emission is first detected at or after peak. However, some of these events may have had emission at or before peak, as most X-ray observations of these sources only commenced as the associated optical/UV transient began to decline. For the other two events that were detected on the rise in the optical/UV, \textit{Swift} XRT observations were taken after peak and showed that PS18kh \citep{holoien18b,velzen19} exhibited weak X-ray emission.
Pre-peak \textit{Swift} observations of ASASSN-18pg only yielded upper limits on the presence of X-ray emission (\citealp{leloudas19}, Holoien et al., \textit{in prep.}).

There are too few counts in the XRT detections to extract a spectrum, but we could determine the hardness ratio, HR=(H$-$S)/(H$+$S), where S is the number of counts in the soft 0.3-2.0 keV energy band and H is the number of counts in the 2.0-10.0 keV energy band. A source is considered soft if it has a $HR=-1$, while it is considered hard if it has a $HR=1$. Interestingly, we find an average hardness ratio of  $\hbox{HR} \sim-0.2$, which is much harder than found for non-jetted, thermal TDEs like ASASSN-14li and ASASSN-15oi which have $\hbox{HR}\sim-0.7$, and more consistent with those seen from the jetted TDEs Swift J1644+57 and Swift J2058-05 \citep{auchettl17, holoien18a}.
	
\begin{figure}
\includegraphics[width=0.9\columnwidth]{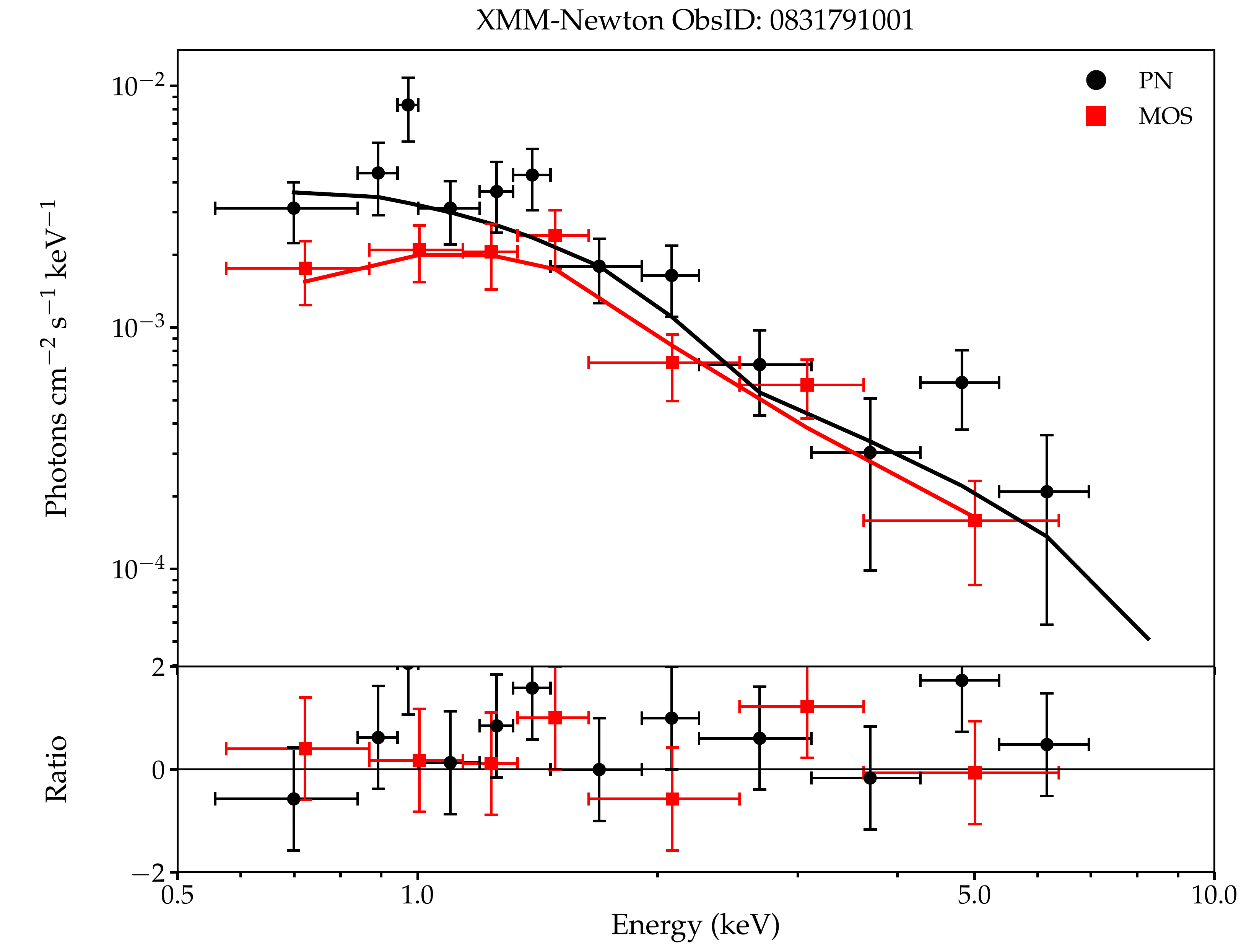}
\includegraphics[width=0.9\columnwidth]{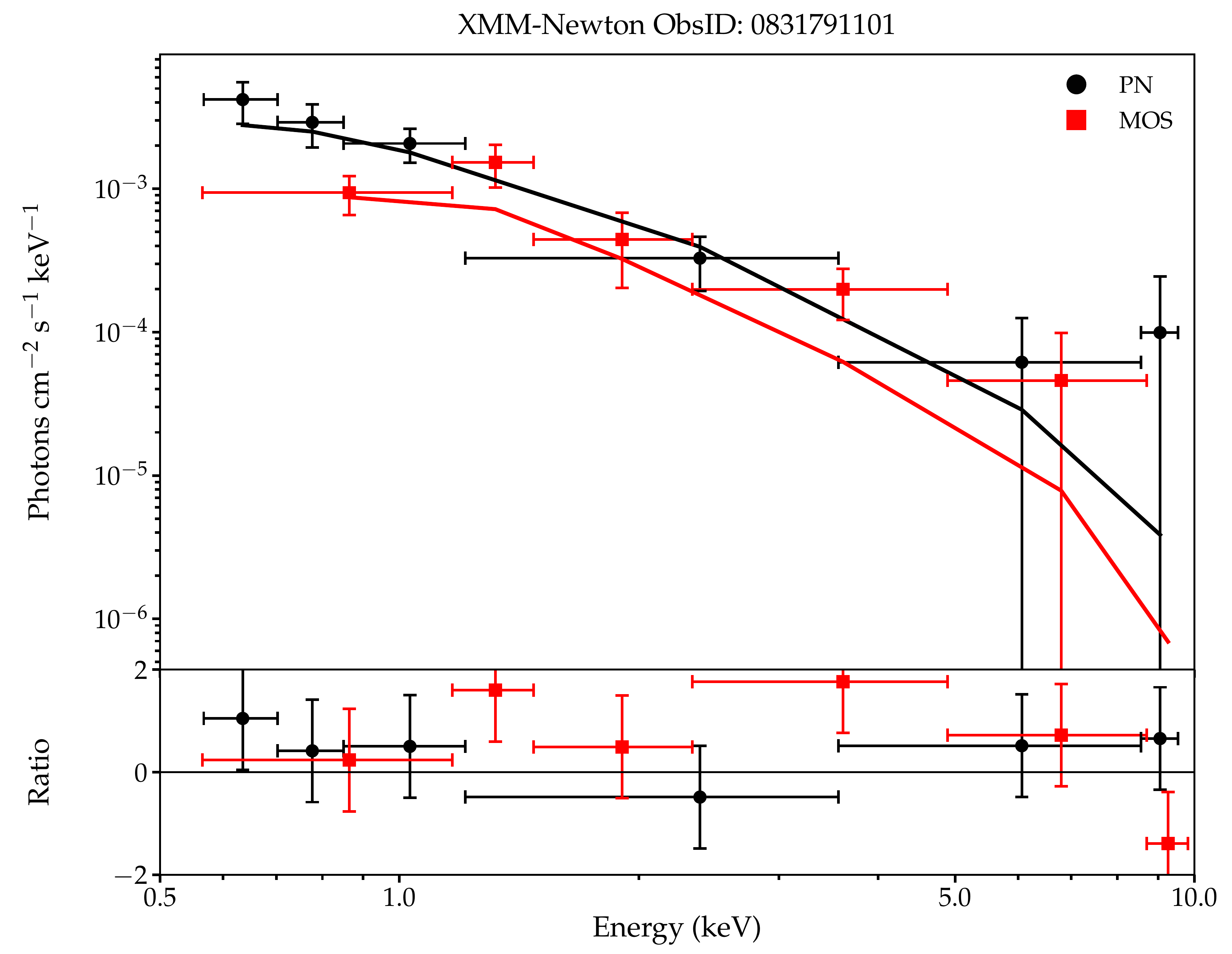}
\caption{\textit{XMM-Newton} PN (black) and merged MOS1+MOS2 (red) spectra of ASASSN-19bt taken $\sim4$ days before peak (top panel) and $\sim42$ days after peak (bottom panel). Each spectrum is binned to have 10 counts per energy bin. (a) The emission from ASASSN-19bt before peak is well fit by an absorbed powerlaw with a photon index of $\Gamma=1.48\pm0.3$, while after peak the emission is well fit with an absorbed powerlaw with a photon index of $\Gamma=2.34^{+0.8}_{-0.6}$. The residuals from the fit are shown in the bottom panel of each figure.}
\label{fig:xmm}
\end{figure}

Due to the hardness of this X-ray emission,  we triggered two deep \textit{XMM-Newton} TOO observation of the source to better constrain its nature. The first observation was taken $\sim$ 4 days before peak, while the second observation was take $\sim$ 42 days after peak and in both observations the source is significantly-detected.  The shallower \textit{Swift} observations taken around this time were only able to determine upper limits due to the faintness of this emission. In Figure \ref{fig:xmm}, we show the resulting PN and MOS1+MOS2 spectra. In the first observation (Figure \ref{fig:xmm} top panel), the spectra are well fit by an absorbed power law model with a photon index of $\Gamma=1.47\pm0.3$ and the Galactic H\textsc{i} column density in the direction of ASASSN-19bt \citep{kalberla05}. Letting the column density ($N_{H}$) vary does not significantly improve the fit. This photon index is consistent with the best fit photon index for the jetted TDE Swift J1644+57 and Swift J2058+05 at late times \citep[e.g.,][]{burrows11,cenko12b,saxton12b,levan15,auchettl17} and those found from AGN \citep[e.g.,][]{ricci17,auchettl18}, implying that the X-ray emission may indicate the presence of a jet. The X-ray emission of several TDEs is well-modeled by a cool blackbody. A blackbody fit to the spectrum of ASASSN-19bt gives a temperature of $0.48\pm0.1$ keV. This is significantly higher than the temperatures found for ASASSN-14li \citep[e.g.,][]{brown16a,kara18} or ASASSN-15oi \cite[][]{holoien18a}, which have temperatures of $\sim50$ eV. Due to the poor signal-to-noise of the spectra, models combining a blackbody with a powerlaw do not improve significantly on our best fit powerlaw model. In the second observation (Figure \ref{fig:xmm} bottom panel), we find that the X-ray emission has softened considerably after peak and is now well fit by an absorbed power law model with a photon index of $\Gamma=2.34^{+0.8}_{-0.6}$, but still consistent with those found from jetted TDES \citep[e.g.,][]{burrows11,cenko12b,saxton12b,auchettl17}. If we fit the spectra using a blackbody instead, we find a slightly lower temperature of $0.20\pm0.1$ keV compared to the first observation, but still higher than that found for other TDEs. Similar to the first observation, fitting the spectra with a blackbody plus a power law does not significantly improve the fit.

\subsection{Spectroscopic Analysis}
\label{sec:spec_anal}

ASASSN-19bt exhibits several broad emission lines that begin to emerge following the 2019 February 07 spectrum, prior to which there are no discernable features. We identify strong emission from \halpha, \ion{He}{1} 5875\AA, and a broad blue feature that spans the \hbeta, \ion{He}{2}~4686\AA, and H$\gamma$ lines. In later epochs, the broad blue feature can be differentiated into likely H$\beta$ and H$\gamma$ lines, and it is apparent that there is little to no emission from the \ion{He}{2} 4686\AA{} line, which is a rarity among TDEs. There is no sign of emission from any of the nitrogen lines identified by \citet{leloudas19} in any epoch, meaning ASASSN-19bt is not among the subset of nitrogen-rich TDEs. The \halpha{} and \ion{He}{1} 5875\AA{} lines are extremely broad once they begin to emerge, with maximum full width at half maxima of $\textrm{FWHM}_{\halpha}\simeq2.7\times10^4$~km~s$^{-1}$ and $\textrm{FWHM}_{\textrm{He I}}\simeq2.1\times10^4$~km~s$^{-1}$. The lines continue to exhibit similar widths throughout the period of observation and begin to develop significantly non-Gaussian shapes in later epochs.

\begin{figure}
\centering
\subfloat{{\includegraphics[width=0.47\textwidth]{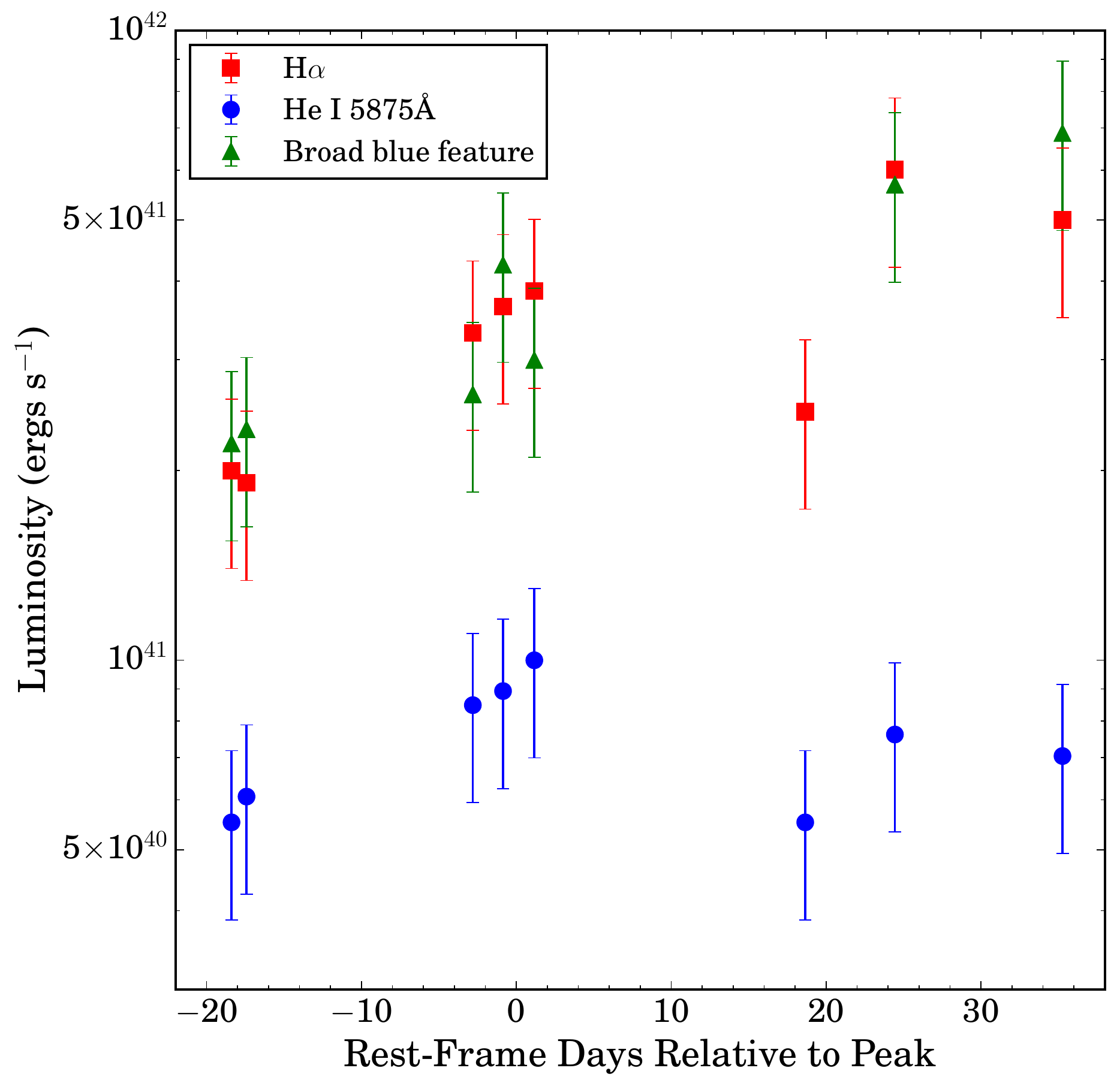}}}
\caption{Evolution of the \halpha{} (red squares) and \ion{He}{2}~5875\AA\ (blue circles) line luminosities and the luminosity of the broad blue emission feature (green triangles) in the spectra of ASASSN-19bt. Errorbars show 30\% errors on the line fluxes. Only epochs where the lines are measurable are shown.}
\label{fig:line_lum}
\end{figure}

We also measured the luminosities of the prominent emission lines in the epochs after the 2019 February 07 spectrum. For each spectrum we subtracted a local continuum estimate for the \halpha{} and \ion{He}{1}~5875\AA{} lines, then measured the integrated line flux using the {\sc IRAF} task {\tt splot}. We also used the same procedure to measure the broad blue feature as a single emission line when possible, as this feature is difficult to break down into its component lines, making it difficult to measure the lines individually. The measured luminosities are given in Table~\ref{tab:line_lum} and are shown in Figure~\ref{fig:line_lum}. Estimating the true error on the line fluxes is difficult given their complex shape, and we assume 30\% errors on the emission fluxes calculated from each epoch.


The broad lines appear roughly 18 days prior to peak and continue to grow stronger throughout the period of observation, with the \halpha{} luminosity roughly tripling between the first detections and the latest spectrum from 2019 April 10. The emergence of spectroscopic emission lines prior to peak light and the continuing strengthening of those lines after peak is similar to what was seen with the TDE ASASSN-18pg \citep{leloudas19}, although in that case the lines were present more than 30 days prior to peak. Increasing line luminosities after peak were also seen in the TDE PS18kh, though in that case the lines were not present until roughly peak light \citep{holoien18a}. ASASSN-19bt thus joins a growing sample of TDEs with early spectra that exhibit increasing line luminosities near peak light, in contrast to early TDE discoveries which were found after peak and which showed lines that became less luminous in the epochs following discovery \citep[e.g.,][]{holoien16a,brown17a}.

The cause for the early, featureless spectra we observe in ASASSN-19bt is not well understood at this time, in part due to the small number of TDEs with spectra obtained this early, but one possible explanation is that at very early times, the material emitting the lines may be moving at very high speeds, as indicated in Figure~\ref{fig:rad_evol2}, which would have the effect of suppressing these features. Further analysis on a larger sample of pre-peak TDE spectra is needed to determine whether other physical effects could play a role in suppressing early line emission, particularly if different TDEs are producing emission via different physical mechanisms, as some of the observations seem to indicate.


\section{Discussion}
\label{sec:disc}

ASASSN-19bt is the first TDE detected by TESS, and, as it happened to fall in the CVZ, it presents us with an unprecedented rising light curve for a TDE. Further, due to its early detection by ASAS-SN only a few days after beginning to brighten, we were able to collect a wide variety of observations prior to peak, including several optical spectra and 10 epochs of \swift{} UVOT and XRT observations. All of these observations combine to make ASASSN-19bt by far the best-observed TDE at early times, and allow us to look at its early evolution in new ways.

While ASASSN-19bt exhibits luminosity, temperature, radius, and spectroscopic evolution all similar to those of other TDEs, it stands out in several ways. Its host appears to be similar to the population of ``shocked post-starburst'' galaxies, a sample of galaxies that are likely to have had a recent episode of star formation, similar to the E$+$A hosts seen for several other TDEs, but which are typically younger and have higher dust obscuration. Our earliest \swift{} observations, obtained more than a month prior to peak, indicate that at very early times the TDE cooled and faded, before the temperature leveled off and the luminosity began to rise again to peak. Based on our data, we can constrain the duration of the early luminosity bump to be $2-15$ rest-frame-days and the duration of the early temperature decline to be at least 12 rest-frame days. This behavior has not been seen in any other TDE, but this may be due to the fact that no TDE has had \swift{} observations at such early epochs.

ASASSN-19bt has weak X-ray emission, and our deep observations with \textit{XMM-Newton} indicate that its X-ray emission is hard compared to that of other thermal TDEs. This may indicate the presence of a jet, as the best-fit photon index is very similar to that found for the jetted TDE Swift J1644+57 and Swift 2058+05 \citep[e.g.,][]{burrows11,cenko12b,saxton12b,auchettl17}. The fact that the emission softens after peak may indicate that prior to peak we are are seeing the stellar debris being expelled in a fast collimated outflow reminiscent of a jet, which then circularises and forms an accretion disk around peak. After peak, the softer photon index may indicate that the hard X-ray emission from the jet is reprocessed by the disk into longer X-ray wavelengths, leading to the shallower photon index, or that the X-ray emission after peak is now dominated by the disk in the form of a soft blackbody component. Due to the poor signal-to-noise of our second XMM-Newton observation we cannot distinguish the jet (powerlaw) and disk (blackbody) components sufficiently. The softening of the X-ray emission seen from ASASSN-19bt after peak is reminiscent of hard and soft states seen in X-ray binaries, where the soft state is dominated by blackbody emission from the disk, while the hard state is dominated by a powerlaw arising from a jet \citep[e.g., see the review by][]{remillard06}.

Similar to what was seen with the TDE PS18kh \citep{holoien18b}, ASASSN-19bt shows no spectroscopic emission features in its earliest spectra. The broad lines only develop as the TDE approaches peak light. Once they appear, we see broad, asymmetric Balmer lines and clearly detect the \ion{He}{1} 5875\AA{} line. Although it is a common feature of TDE spectra, there is no discernible \ion{He}{2} 4686\AA{} line. 

The TESS light curve allows us to place a strong constraint on when the TDE began to brighten, and we find that the time between first light and the peak of its light curve was 41 rest-frame days. The very early rise in the TESS flux is well-fit by a power law that is consistent with the $t^2$ ``fireball'' model found for the early rise of supernovae. However, since the temperature of ASASSN-19bt does not appear to be constant during this time, the consistency with the fireball model is likely coincidental. The TESS light curve and early \swift{} SED fits allow us to trace the size of the emission region of the TDE to an unprecedented early phase, when it had a likely size corresponding to tens of gravitational radii. We are also able to trace the likely velocities based on the radius evolution, finding that it is either expanding at a rate well below the local escape speed or it is a very compact emission region at a much larger distance from the BH than its apparent size.

While no other TDE has a light curve that captures the rise to peak like the TESS light curve of ASASSN-19bt, some TDEs have been observed at $30-40$ days prior to peak---notably, PS18kh \citep{holoien19a,velzen19} and ASASSN-18pg \citep{leloudas19}. It is unclear whether these other objects can be fit with the same power-law model that we use here to fit the rise of ASASSN-19bt, but it is possible that these and other future discoveries will have data close enough to first light to be able to constrain the power-law parameters. Using ASASSN-19bt as a guide, this likely requires data within the first $\sim1-2$ weeks after the light curve begins to rise in order to have enough data points to constrain the fit.

ASASSN-19bt is a poster child for a new era in early TDE studies. While not all TDEs will be found in the TESS CVZ, or be detected by TESS at all, TDEs are being found in greater numbers and at earlier times by surveys like ASAS-SN. This allows us to trigger multiwavelength follow-up observations earlier and probe the emission region at very early times. As more TDEs are found and observed extensively prior to peak, we will be able to build a better theoretical understanding of how the stellar debris evolves and how the accretion disk forms after disruption, hopefully illuminating the physics behind the later emission that we have been limited to studying for the majority of the TDEs found to date.

\acknowledgments

We thank the \swift{} PI, the Observation Duty Scientists, and the science planners for promptly approving and executing our \swift{} observations. We thank the \textit{XMM-Newton} team for promptly scheduling and executing our TOO observations. We thank the Las Cumbres Observatory and its staff for its continuing support of the ASAS-SN project. This research utilizes Las Cumbres Observatory observations obtained with time allocated through the National Optical Astronomy Observatory TAC  (NOAO Prop. ID 2018B-0110, PI: P. Vallely).

ASAS-SN is supported by the Gordon and Betty Moore Foundation through grant GBMF5490 to the Ohio State University and NSF grant AST-1515927. Development of ASAS-SN has been supported by NSF grant AST-0908816, the Mt. Cuba Astronomical Foundation, the Center for Cosmology and AstroParticle Physics at the Ohio State University, the Chinese Academy of Sciences South America Center for Astronomy (CASSACA), the Villum Foundation, and George Skestos.

PJV is supported by the National Science Foundation Graduate Research Fellowship Program Under Grant No. DGE-1343012. KAA is supported by the Danish National Research Foundation (DNRF132). KZS, CSK, and TAT are supported by NSF grants AST-1515876, AST-1515927, and AST-1814440. CSK is also supported by a fellowship from the Radcliffe Institute for Advanced Studies at Harvard University. KDF is supported by Hubble Fellowship grant HST-HF2-51391.001-A, provided by NASA through a grant from the Space Telescope Science Institute, which is operated by the Association of Universities for Research in Astronomy, Incorporated, under NASA contract NAS5-26555. Support for JLP is provided in part by FONDECYT through the grant 1191038 and by the Ministry of Economy, Development, and Tourism's Millennium Science Initiative through grant IC120009, awarded to The Millennium Institute of Astrophysics, MAS. SD acknowledges Project 11573003 supported by NSFC. TAT acknowledges support from a Simons Foundation Fellowship and from an IBM Einstein Fellowship from the Institute for Advanced Study, Princeton. Support for this work was provided by NASA through Hubble Fellowship grant \#51386.01 awarded to RLB by the Space Telescope Science Institute, which is operated by the Association of  Universities for Research in Astronomy, Inc., for NASA, under contract NAS 5-26555. JS acknowledges support from the Packard Foundation.

This paper includes data collected by the TESS mission, which are publicly available from the Mikulski Archive for Space Telescopes (MAST). Funding for the TESS mission is provided by NASA's Science Mission directorate.

This paper includes data gathered with the 6.5-meter Magellan Telescopes located at Las Campanas Observatory, Chile.

Based on observations obtained at the Southern Astrophysical Research (SOAR) telescope, which is a joint project of the Minist\'{e}rio da Ci\^{e}ncia, Tecnologia, Inova\c{c}\~{o}es e Comunica\c{c}\~{o}es (MCTIC) do Brasil, the U.S. National Optical Astronomy Observatory (NOAO), the University of North Carolina at Chapel Hill (UNC), and Michigan State University (MSU).

\software{FAST (Kriek et al. 2009), IRAF (Tody 1986, Tody 1993), HEAsoft (Arnaud 1996), XSPEC (v12.9.1; Arnaud 1996)}

\bibliography{bibliography.bib}
\bibliographystyle{apj}


\begin{deluxetable}{cccccc}[h!]
\tabletypesize{\footnotesize}
\tablecaption{Spectroscopic Observations of ASASSN-19bt}
\tablehead{
\colhead{Date} &
\colhead{Telescope} &
\colhead{Instrument} &
\colhead{Grating} &
\colhead{Slit} &
\colhead{Exposure Time}}
\startdata
2019 January 31.20 & Magellan Clay 6.5-m & LDSS-3 & VPH-All & 1\farcs{00} blue & 2x600s \\
2019 February 01.26 & du Pont 100-inch & WFCCD & Blue & 1\farcs{65} & 3x900s \\
2019 February 07.03 & Magellan Baade 6.5-m & IMACS f/2 & 400 l/mm & 1\farcs{20} & 1x300s \\
2019 February 14.04 & Magellan Baade 6.5-m & IMACS f/2 & 300 l/mm & 0\farcs{90} & 2x600s \\
2019 February 15.03 & Magellan Baade 6.5-m & IMACS f/2 & 300 l/mm & 0\farcs{90} & 2x600s \\
2019 March 02.02 & du Pont 100-inch & WFCCD & Blue & 1\farcs{65} & 3x600s \\
2019 March 04.03 & du Pont 100-inch & WFCCD & Blue & 1\farcs{65} & 3x600s \\
2019 March 06.10 & du Pont 100-inch & WFCCD & Blue & 1\farcs{65} & 3x600s \\
2019 March 24.05 & SOAR 4.1-m & Goodman & 400 l/mm & 0\farcs{95} & 1x1200s \\
2019 March 29.00 & Magellan Clay 6.5-m & LDSS-3 & VPH-All & 1\farcs{00} blue & 4x600s \\
2019 April 10.20 & du Pont 100-inch & WFCCD & Blue & 1\farcs{65} & 3x600s \\
\enddata 
\tablecomments{Date, telescope, instrument, grating, slit size, and exposure time for each of the spectroscopic observations obtained of ASASSN-19bt for the initial classification of the transient and as part of our follow-up campaign.} 
\label{tab:spec_details} 
\end{deluxetable}

\begin{deluxetable}{cccc}[h!]
\tabletypesize{\footnotesize}
\tablecaption{X-ray Luminosities (0.3-10.0 keV)}
\tablehead{
\colhead{Observation} &
\colhead{MJD} &
\colhead{Rest-Frame Days Relative to Peak} &
\colhead{X-Ray Luminosity}}
\startdata
{\swift} 001$-$005 & 58521.4 & $-24.9$ & $<7.68\times10^{40}$ \\
{\swift} 006$-$008 & 58535.7 & $-10.9$ & $6.73^{+3.57}_{-3.61}\times10^{40}$ \\
XMM 0831791001 & 58543.2 & $-3.6$ & $4.48^{+0.77}_{-0.78}\times10^{40}$ \\
{\swift} 009$-$013 & 58550.0 & $3.0$ & $<6.62\times10^{40}$ \\
{\swift} 014$-$016 & 58569.6 & $22.1$ & $<1.26\times10^{41}$ \\
{\swift} 017$-$019 & 58581.4 & $33.6$ & $<8.69\times10^{40}$ \\
XMM 0831791101 & 58589.0 & $-42.1$ & $1.24^{+0.40}_{-0.38}\times10^{40}$ \\
\enddata 
\tablecomments{X-ray luminosities and $3\sigma$ upper limits on the X-ray luminosity from our {\swift} XRT and \textit{XMM-Newton} observations. {\swift} data were binned in time to increase the signal-to-noise of the observations, and the epochs combined for each bin are indicated in Column 1.}
\label{tab:xray} 
\end{deluxetable}

\begin{deluxetable}{cccc}[h!]
\tabletypesize{\footnotesize}
\tablecaption{Measured Line Luminosities}
\tablehead{
\colhead{Rest-Frame Days Relative to Peak} &
\colhead{\halpha{} Luminosity} &
\colhead{\ion{He}{1} Luminosity} &
\colhead{Broad Blue Feature Luminosity}}
\startdata
-18.39 & ($2.00\pm0.60$)$\times10^{41}$ & ($0.55\pm0.17$)$\times10^{41}$ & ($2.21\pm0.66$)$\times10^{41}$ \\ 
-17.42 & ($1.91\pm0.57$)$\times10^{41}$ & ($0.61\pm0.18$)$\times10^{41}$ & ($2.33\pm0.70$)$\times10^{41}$ \\ 
-2.81 & ($3.31\pm0.99$)$\times10^{41}$ & ($0.85\pm0.25$)$\times10^{41}$ & ($2.64\pm0.79$)$\times10^{41}$ \\ 
-0.86 & ($3.64\pm1.09$)$\times10^{41}$ & ($0.89\pm0.27$)$\times10^{41}$ & ($4.25\pm1.27$)$\times10^{41}$ \\ 
1.16 & ($3.86\pm1.16$)$\times10^{41}$ & ($1.00\pm0.30$)$\times10^{41}$ & ($3.00\pm0.90$)$\times10^{41}$ \\ 
18.65 & ($2.48\pm0.74$)$\times10^{41}$ & ($0.55\pm0.17$)$\times10^{41}$ & --- \\ 
24.44 & ($6.01\pm1.80$)$\times10^{41}$ & ($0.76\pm0.23$)$\times10^{41}$ & ($5.69\pm1.71$)$\times10^{41}$ \\ 
35.26 & ($5.00\pm1.50$)$\times10^{41}$ & ($0.7\pm0.21$)$\times10^{41}$ & ($6.87\pm2.06$)$\times10^{41}$ \\ 
\enddata 
\tablecomments{Line luminosities of the \halpha{} and \ion{He}{1} 5875\AA{} lines and the broad blue feature spanning \hbeta, \ion{He}{2} 4686\AA, and H$\gamma$ measured from the follow-up spectra of ASASSN-19bt. All luminosities are quoted in erg~s$^{-1}$. The broad blue feature was not measurable in all epochs, and no lines were measurable in the spectra taken prior to 2019 February 12. The uncertainties are 30\% uncertainties on the measured fluxes.} 
\label{tab:line_lum} 
\end{deluxetable}

\end{document}